\documentstyle[aaspp4]{article}

\lefthead{MENOU, NARAYAN \& LASOTA}
\righthead{A Population of Faint Non-Transient LMBHBs}

\def\spose#1{\hbox to 0pt{#1\hss}}
\def\lta{\mathrel{\spose{\lower 3pt\hbox{$\mathchar"218$}}
     \raise 2.0pt\hbox{$\mathchar"13C$}}}
\def\gta{\mathrel{\spose{\lower 3pt\hbox{$\mathchar"218$}}
     \raise 2.0pt\hbox{$\mathchar"13E$}}}

\begin{document}

\def\Journal#1#2#3#4{#4, {#1} {#2}, #3}
\def\MNRAS{{MNRAS}}
\def\APJ{{ApJ}}
\def\AAP{{A\&A}}
\def\NAT{{Nature}}
\def\ASTRPH{{Astro-ph/}}
\def\ARAA{{ARAA}}
\def\APJL{{ApJ Lett.}}
\def\PASJ{{PASJ}}

\title{A Population of Faint Non-Transient Low Mass Black Hole Binaries}

\author{Kristen Menou\altaffilmark{1} and  Ramesh Narayan}
\affil{Harvard-Smithsonian Center for Astrophysics, 60 Garden Street,\\ 
Cambridge, MA 02138, USA.\\ kmenou@cfa.harvard.edu, rnarayan@cfa.harvard.edu}
\and
\author{Jean-Pierre Lasota}
\affil{UPR 176 du CNRS, D\'epartement d'Astrophysique Relativiste et de
Cosmologie, Observatoire de Paris, Section de Meudon, 92195 Meudon C\'edex,
France.\\ lasota@obspm.fr}
\altaffiltext{1}{also UPR 176 du CNRS, D\'epartement d'Astrophysique Relativiste et de
Cosmologie, Observatoire de Paris, Section de Meudon, 92195 Meudon C\'edex,
France.}

\begin{abstract}
We study the thermal and viscous stability of accretion flows in Low
Mass Black Hole Binaries (LMBHBs). We consider a model in which an
inner advection-dominated accretion flow (ADAF) is surrounded by a
geometrically thin accretion disk, the transition between the two zones
occurring at a radius $R_{\rm tr}$. In all the known LMBHBs, $R_{\rm
tr}$ appears to be such that the outer disks could suffer from a global
thermal-viscous instability.  This instability is likely to
cause the transient behavior of these systems.  However, in
most cases, if $R_{\rm tr}$ were slightly larger than the estimated
values, the systems would be globally stable.  This suggests that a
population of faint persistent LMBHBs with globally stable outer disks
could be present in the Galaxy.  Such LMBHBs would be hard to
detect because they would lack large amplitude outbursts, and because
their ADAF zones would have very low radiative efficiencies, making the
systems very dim.  
We present model spectra of such systems covering the optical and
X-ray bands.
\end{abstract}


\section{Introduction}

All known Low Mass Black Hole Binaries (LMBHBs) and several Low Mass
Neutron Star Binaries (LMNSBs) are transient (see e.g.  Tanaka \&
Lewin 1995, van Paradijs \& McClintock 1995, White, Nagase \& Parmar 1995).  
These systems are also
known as `Soft X-ray Transients' or SXTs.  In both LMBHBs and LMNSBs,
a low mass secondary star (main sequence or sub-giant) transfers mass
by Roche-lobe overflow to the compact primary. The transfered matter
flows on to the primary via an accretion disk.

For effective temperatures between about 5000~K and 8000~K, accretion
disks described by the `$\alpha$ - viscosity' prescription are
thermally and viscously unstable due to large changes in the opacity
as a result of hydrogen recombination.  This instability is believed
to trigger outbursts of dwarf novae.  A dwarf nova is a cataclysmic
binary (CB) system (Warner 1995) in which a white dwarf accretes from
a low-mass companion at a mass accretion rate such that a steady disk
would, at least over some range of radius, have an effective
temperature in the thermally unstable range 5000--8000~K.  Since a
steady disk is not allowed, the system undergoes a limit cycle where
for some of the time (i.e during outburst) the mass accretion rate is
high and the disk is entirely in a hot state with $T_{\rm eff} \gta
8000$~K, and for the rest of the time (during quiescence) the
accretion rate is low and the entire disk is in a cold state with
$T_{\rm eff} \lta 5000$~K.

The disk instability model (DIM; see Cannizzo 1993 for a review and
Hameury et al.  1998 for a recent version of the model) describes well
the principal properties of dwarf novae.  However, in its simplest
version, the model cannot account for the variety of observed
behaviors (e.g. Lasota \& Hameury 1998). In particular, the `standard' version
of the DIM cannot account for the properties of WZ Sge-type dwarf novae with
large outburst amplitudes and long recurrence times.  (Some of these
systems are also called TOADs, but we will not use this name here.)

The WZ Sge-type dwarf novae resemble transient LMBHBs and LMNSBs
(Lasota 1996a,b; Kuulkers, Howell \& van Paradijs 1996) and it was
suggested quite early (van Paradijs \& Verbunt 1984; Cannizzo, 
Ghosh \& Wheeler 1995; Huang \& Wheeler 1989; Mineshige \&
Wheeler 1989) that the same mechanism that triggers outbursts in dwarf
novae also operates in transient LMBHBs and LMNSBs.  However, detailed
attempts to explain the properties of SXT outbursts (Mineshige \&
Wheeler 1989) failed to reproduce the observed long recurrence times,
which are of the order of 20 to 50 years in transient
LMBHBs. Moreover, spectral observations of quiescent SXTs disagree
seriously with the predictions of the DIM (Narayan, McClintock \& Yi
1996; Lasota 1996a).  In particular, the observed X-ray emission in
LMBHBs is more than four orders of magnitude larger than that
predicted by the DIM. Not surprisingly similar problems
are encountered in the dwarf nova case (Lasota 1996b).

Narayan et al. (1996) showed that the quiescent disk properties in
SXTs could be explained with a model in which the accretion flow
consists of two components: an outer CB-type thin accretion disk and
an inner, advection-dominated accretion flow (ADAF) in which the
radiative efficiency is very low.  A similar model was proposed by
Lasota, Hameury \& Hur\'e (1995) for WZ Sge. There, however, the inner
`hole' in the CB accretion disk could be due to a weak magnetic field
(Patterson et al. 1997).  In the Narayan et al. (1996) model quiescent
optical and UV radiation was emitted by the outer disk and the X-ray
radiation came from the ADAF.  The model, however, had a serious
problem (Lasota, Narayan \& Yi 1996); some regions of the outer disk
had effective temperatures within the unstable range, $5000$~K
$<T_{\rm eff}<8000$~K (Wheeler 1996).

A modified version of the two-component model was subsequently proposed
by Narayan, Barret \& McClintock (1997) which eliminates this problem.
The transition radius is larger so that the outer disk 
is cool enough to be stable at all radii ($T_{\rm eff}<5000$~K).  
In this model, the optical and UV emission are from the
ADAF and not the outer disk. The ADAF
also produces the X-ray emission. The outer disk contributes only in
the infrared, and even in this band its emission is overshadowed by the
radiation from the secondary star.

It was noticed by Lasota et al. (1996) that the transition
radii of SXTs, as determined from H$\alpha$ line observations, are
close to the values for which the outer thin disk would be {\it
globally} stable.
(By global stability, we mean that the disk can stably support a constant 
mass accretion rate; thus, the system will not undergo a limit cycle.)
Based on this
fact, Lasota et al. (1996) suggested that there may exist other closely
related systems with a slightly larger transition radius in which the
outer disk would, in fact, be globally stable.  
In the absence of significant variations of the mass-transfer
rate such systems would not
have transient outbursts, but would be steady and persistent.  In
addition, they would be extremely dim in X-rays, and quite
under-luminous even in the optical, so that they would be very
difficult to detect.  This `hidden population' is the main subject of
this paper.

For the remainder of the paper, we refer to optically thin
two-temperature ADAFs simply as ADAFs (see Chen et al. 1995 for a
distinction between various branches of accretion solutions). We refer
to LMBHBs with a globally stable truncated thin disk (the topic of
this paper) as faint persistent LMBHBs, or equivalently cold steady
systems, and refer to the more standard transient LMBHBs with a globally 
unstable truncated thin disk as transient LMBHBs, or equivalently BH
SXTs.  The symbols $M$, $\dot{M}$ and $R$ refer to mass, mass
accretion rate and radius in physical units, respectively, while the
symbols $m$, $\dot{m}$ and $r$ refer to the same quantities in solar
units ($m=M/M_{\odot}$), in Eddindgton units
($\dot{m}=\dot{M}/\dot{M}_{\rm Edd}$, $\dot{M}_{\rm Edd} = 1.39 \times
10^{18}~m~{\rm g~s^{-1} }$) and in Schwarzschild units ($r=R/R_S$,
$R_S = 2.95 \times 10^5~m~{\rm cm}$).

\section{The Global Stability of LMBHB Accretion Flows}

We discuss the structure and stability of LMBHB accretion flows
according to the model of quiescent SXTs developed by Narayan et al.
(1997a).  The accretion occurs primarily as a thin disk from the outer
radius $r_{\rm out}$ down to a transition radius $r_{\rm tr}$, and
then switches to a two-temperature ADAF for $r<r_{tr}$.  The
transition itself is not explained from first principles, but is
introduced empirically.

This two-zone model successfully explains the observed properties of
several SXTs in quiescence (Narayan et al. 1996, 1997a; Hameury et
al. 1997b).  Esin et al.  (1997, 1998) show that such models also
explain the various spectral states of black hole candidates, such as
the high state, the intermediate state, and the low state, as
sequences of varying accretion rate and location of the transition
radius.

Lasota et al. (1996) discussed the mechanism for the origin of
outbursts in BH SXTs and argued that the outbursts must be triggered by
an instability in the outer thin disk component of the flow rather than
in the inner ADAF zone.  Two recent studies have strengthened the
argument.  First, Hameury et al. (1997b) have successfully reproduced
all the properties of the rise to outburst of the BH SXT GRO
J1655-40 with a model in which the outburst is triggered by the
thermal-viscous instability of the outer disk.  Second, detailed
stability studies have confirmed that two-temperature ADAFs are
essentially thermally and viscously stable (Kato et al. 1996, 1997; 
Wu, 1997\footnote{Note that 
Wu (1997) used vertically averaged 
equations to study the stability of the flow. Kato et al. (1997) have
argued that this is incorrect.}). 
We therefore focus in this paper on the stability
properties of the outer disk.

In the DIM, the stability criterion for a disk annulus can be
expressed in terms of the local mass accretion rate $\dot{M}(R)$. If
$\dot{M}(R)$ lies between two critical values, $\dot{M}_{\rm
crit}^+(R)$ and $\dot{M}_{\rm crit}^-(R)$, then the effective temperature 
of the disk lies in the unstable range (5000-8000~K), 
and the local annulus is thermally and viscously unstable.

The disk is globally stable, and does not experience outbursts,
whenever all its annuli are stable.  This can be achieved in two ways.
First, the disk can be in a hot, stable equilibrium if $\dot{M}(R)=$
constant $> \dot{M}_{\rm crit}^+(R)$ for all radii $R$ in the disk.
This is the standard scenario for persistent X-ray binaries and has
been discussed 
recently by van Paradijs (1996), who included the effect of X-ray
irradiation (but see also Dubus et al. 1998).  Second, the disk can be
in a cold, stable equilibrium if $\dot{M}(R)=$ constant $<\dot{M}_{\rm
crit}^-(R)$ at all $R$.  It is the second option that we are interested
in here.

Calculations of thin disk vertical equilibria (\cite{hmdlh98}; see
also Ludwig, Meyer-Hofmeister \& Ritter 1994) give
\begin{equation}
\dot{M}_{\rm crit}^-(R) = 4.0 \times 10^{15} ~ \alpha^{-0.04} {m_1}^{-0.89} 
\left( {R \over 10^{10} \; \rm cm} \right)^{2.67}
~\rm g~s^{-1},
\label{eq:mdcrit-}
\end{equation}
where $\alpha \leq 1$ is the turbulent viscosity parameter (Shakura \&
Sunyaev 1973), $m_1$ is the mass of the accreting primary in solar
mass units, and $R$ is the radius of the annulus under consideration.
Since $\dot{M}_{\rm crit}^-(R)$ is a steep function of $R$, we are
unlikely to find a system with a cold, stable equilibrium if the thin
accretion disk extends all the way down to the last stable orbit
around a black hole.  The limit on $\dot M$ in this case is so low (of
order $10^6~{\rm g~s^{-1}}$ typically) that any system that satisfies
the constraint $\dot{M} < \dot{M}_{\rm crit}^-(R_{\rm in})$ would be
observationally very uninteresting (this statement is true even if we
use very low values of $\alpha$, e.g. $10^{-4}$).  However, if the
disk is truncated at a radius that is significantly outside the
marginally stable orbit, then it is possible to have a globally stable
cold flow with reasonable values of $\dot M$.

Eq.~(\ref{eq:mdcrit-}) shows that for a given, constant, accretion
rate $\dot M$ there exists a critical radius $R_{\rm crit}$ given by

\begin{equation}
R_{\rm crit}  =  6\times 10^{9}
\left({\dot{M}\over 10^{15} \rm g~s^{-1}}\right)^{0.375} \alpha^{0.015}
{m_1}^{0.33}  \; \rm cm.
\label{eq:recrit}
\end{equation}
If the transition from the outer disk to the ADAF occurs at a 
radius $R_{\rm tr}$ greater than this
$R_{\rm crit}$, the flow will be cold and globally stable. 

In two-component accretion flows, the transition between the thin disk
and the ADAF most likely occurs over a range of radius by the gradual
evaporation of the thin disk material (Meyer \& Meyer-Hofmeister 1994;
Narayan \& Yi 1995).  The transition is thus not abrupt, although the
width of the transition region is likely to be narrow (Esin et
al. 1997; \cite{ail98}).  We assume for the remainder of this work
that the evaporation process does not affect the disk stability.  This
is a reasonable approximation if the evaporation law is steep (for
example, the $1/r$ law for the coronal accretion rate used by Esin et
al. 1997) since the accretion rate in the thin disk is then almost
unchanged except very close to the transition radius.  Note that
evaporation reduces the accretion rate in the disk. According to
Eq.~(\ref{eq:mdcrit-}), it should therefore increase the stability of
the disk.

\subsection{Prescription for the Transition Radius R$_{\rm tr}$}

The location of $R_{\rm tr}$ is ultimately determined by the
physics of the evaporation of gas from the thin accretion disk to the ADAF.
This physics is poorly understood.  Several mechanisms have
been proposed in the literature (Meyer \& Meyer-Hofmeister 1994,
Narayan \& Yi 1995, Honma 1996) to explain the transition, but none of
these models has reached a sufficient level of detail to allow
quantitative predictions (see e.g. Narayan, Mahadevan \& Quataert 1998
for a discussion).

Quite independent of the actual process of evaporation, various limits 
can be derived on the value of $R_{\rm tr}$.  First,
the dynamics of the incoming stream from the secondary 
leads to a maximum value of $R_{\rm tr}$.  At the same
time, observations of the $H_{\alpha}$ line from the disk provide an
independent limit on $R_{\rm tr}$.  These two constraints are
consistent with each other, as we show in \S2.1.1 and \S2.1.2.  Another
limit arises from the fact that for a given mass accretion rate,
an ADAF is not allowed beyond a certain maximum radius $R_{\rm max}$.  
We estimate this limit in \S2.1.3.

These arguments give only upper limits on $R_{\rm tr}$.  In this paper
we assume that the ADAF extends out to the largest radius it is
allowed (Narayan \& Yi 1995), that is to the smallest of the various
upper limits. Observations of quiescent SXTs are consistent with this
assumption, as we discuss in \S2.1.4, and we assume that the systems
we discuss in this paper also satisfy this rule. However, we note that
the assumption is apparently not satisfied in the nucleus of NGC 4258
(Gammie, Blandford \& Narayan 1998).

\subsubsection{Constraint from Stream Dynamics}

According to the calculations of Lubow \& Shu (1976) and Lubow (1989),
a substantial fraction of the stream of matter from the secondary
overflows the rim of the accretion disk and impacts at a radius
$R_{\rm impact}$ that is significantly smaller than the outer radius
of the disk, $R_{\rm out}$.  Such overflows are seen in `dipping' low
mass X-ray binaries (see e.g. Uno et al. 1997; Frank, King \&
Lasota 1987) and in cataclysmic variables (see Warner 1995).

We do not expect the ADAF zone to extend to radii larger than the
impact radius of the stream. This is because the
cold stream moves very rapidly (essentially at free-fall speed) from the 
Roche lobe of the secondary to the impact point; it is unlikely that the 
material in the stream could evaporate into an ADAF in this short time.
At the very least,  
we expect the cold incoming stream to form an annulus at the impact radius 
$R_{\rm impact}$. More likely, the material will spread a little under 
the influence of viscous evolution 
before it evaporates fully. Thus we expect $R_{\rm tr} \lta R_{\rm impact}$.

Lubow's calculations show that, in the free-stream approximation, all
overflowing streams converge to a single impact radius which scales to
a good approximation as $R_{\rm impact} \simeq 0.48 R_{\rm
circ}$ (the coefficient corresponds to a low mass ratio $q=m_2/m_1<
0.5$, which is valid for all the systems we consider).  Here, $R_{\rm
circ}$ is the circularization radius of the binary (Frank et al.
1992),
\begin{eqnarray} 
\frac{R_{\rm circ}}{a} & = & (1+q)[0.5-0.227 \times \log q]^4,\\ 
a & = & 3.5 \times 10^{10}~m_1^{1/3}~(1+q)^{1/3}~P_{\rm hr}^{2/3}~{\rm
cm},
\label{eq:rcirc} 
\end{eqnarray} 
where $a$ is the orbital separation of the binary, and $P_{\rm hr}$ is
the orbital period in hours.  Since we have argued that $R_{\rm impact}$ is 
an upper limit on the transition radius, we write
\begin{equation}
R_{\rm tr}=f_t \ R_{\rm circ},\qquad f_t<0.48.
\label{eq:rtr}
\end{equation}

The systems that we consider span almost two orders of magnitude in
orbital period.  Since the circularization radius varies as $P_{\rm
orb}^{2/3}$, the impact radius also spans a wide range.  In principle,
the parameter $f_t$ may vary in some complicated fashion as a function
of $R_{\rm impact}$.  For simplicity, we treat $f_t$ as a free
parameter, which absorbs all the complicated physics associated with
the stream dynamics, and consider values between 0.1 and 0.48.

\subsubsection{Constraint from $H_\alpha$ Emission Line Width}

\begin{figure}
\plotone{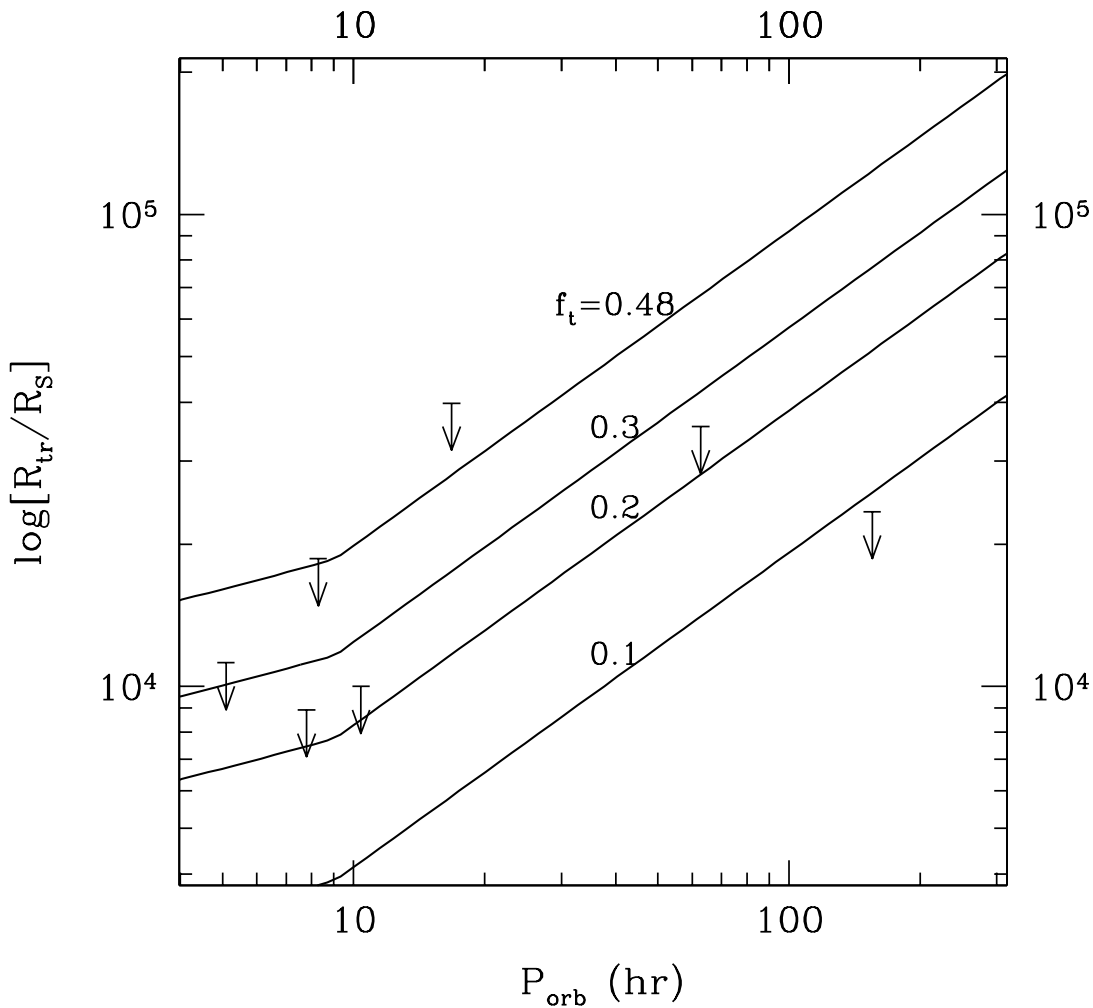}
\caption{Shows observational upper limits (downward arrows) on the
transition radius $R_{\rm tr}$ in the seven quiescent BH SXTs listed
in Table~1.  The limits were obtained from the widths of the $H_{\rm
\alpha}$ emission line in the quiescent spectra of the systems.  Solid
lines show the predicted variation of $R_{\rm tr}$ with orbital
period, based on stream dynamics, for a `typical' LMBHB with $m_1=7$
and $m_2=\min[0.11 P_{\rm hr},1]$.  Four cases are shown, with
$f_t \equiv R_{\rm tr}/R_{\rm circ}= 0.48,~0.3,~0.2,~0.1$.  The $H_\alpha$
line limits require that $f_t \lta 0.25$, except in the long period
system V404 Cyg which requires $f_t \lta 0.1$. \label{fig:halpha}}
\end{figure}

Observations of $H_{\alpha}$ emission lines from the disk 
constrain the value of $R_{\rm tr}$ in quiescent BH SXTs.  
As emphasized by Narayan et al. (1996), half the full-width at zero-intensity
of the $H_\alpha$ emission line provides a good lower limit to the 
speed of matter, $v_{\rm tr}$,  at the transition radius.
This can be translated into an
upper limit on the transition radius through the relation (assuming a
Keplerian disk)
\begin{equation}
r_{\rm tr}<\frac{1}{2} \left( \frac{c  \sin i}{v_{\rm tr}} 
\right)^2,
\label{eq:rin}
\end{equation}
where $c$ is the speed of light and $i$ is the inclination of the
binary system to the line of sight.  Narayan et al. (1996) used this
relation to derive upper limits on $r_{\rm tr}$ in the SXTs
A0620-00, V404 Cyg and Nova Mus 91, and Hameury et al. (1997b) did
the same for GRO J1655-40.  We derive upper limits for three other
quiescent systems, GRO J0422+32, GS2000+25 and Nova Oph 77,
using the inclinations summarized in Table~\ref{tab:systems} and the
following lower limits on $v_{\rm tr}$, collected from the
literature: $v_{\rm tr} > 1000~{\rm km \ s^{-1}}$ in GRO J0422+32
(Garcia et al. 1996, Orosz \& Bailyn 1995), $v_{\rm tr} > 1400 \ {\rm
km \ s^{-1}}$ in GS2000+25 (Harlaftis et al. 1996) and $v_{\rm tr} >
1000~{\rm km \ s^{-1}}$ in Nova Oph. 77 (Harlaftis et al.  1997).
The upper limits on $r_{\rm tr}$ for all seven systems are
indicated by downward arrows in Fig.~\ref{fig:halpha}.

Figure~\ref{fig:halpha} also shows the predictions for the transition
radius according to Eq.~(\ref{eq:rtr}).  Four values of the
parameter $f_t$ are considered: $f_t=$0.1, 0.2,
0.3, 0.48.  Note that the circularization radius and hence the impact
radius in a binary system depends on the masses of the stars.
Therefore, a knowledge of the masses as functions of the orbital
period of the system is required in order to use
Eq.~(\ref{eq:rtr}).  The mass of the secondary at a given period is
theoretically constrained by binary evolution theory. Semi-detached
binaries in which main-sequence donors fill their Roche-lobes
satisfy the relation $m_2 \approx 0.11 P_{\rm hr}$, where $P_{\rm hr}$
is the orbital period of the system in hours (e.g. Frank et al. 1992).
When the secondary is a subgiant or a giant, the period mass relation
is less simple because it depends also on the core mass of the star
(e.g. King 1988).  For simplicity, in Fig.~\ref{fig:halpha} we have
assumed that we have `typical' systems with $m_1=7$ and $m_2=\min [0.11
P_{\rm hr},1]$.  This assumption is valid for four of the SXTs
listed in Table~\ref{tab:systems}.  The systems J0422+32, J1655-40 and
V404 Cyg deviate somewhat from this assumption.  We have derived
values of $R_{\rm tr}$ for these systems, taking into account their
actual masses; the results are not very different.

Technically, the $H_\alpha$ line widths only give upper limits on the 
transition radius; the fact that $H_\alpha$ emission is not observed from
gas orbiting at higher speeds may merely mean that such high velocity 
material does not emit enough radiation to be detectable 
(small emitting area). We make the stronger assumption that the
high velocity material is absent altogether in the cold disk, 
so that the 
$H_\alpha$ line widths directly give an estimate of $R_{\rm tr}$ to within
a factor of a few.
In that case, we see that most of the systems shown in 
Fig.~\ref{fig:halpha} are consistent with $f_t
\sim 0.25$.  More precise calculations of the circularization
radii of J0422+32, J1655-40 and V404 Cyg, taking into account $m_1$ and
$m_2$ in Table~\ref{tab:systems}, show that the first two of these
systems are also in general agreement with the condition $f_t\sim0.25$.
The only peculiar object is V404 Cyg, which appears to require
$f_t<0.1$.  This system, however, has the longest orbital period and
may behave differently because of this (see \S 2.1.3). 
\.Zycki, Done \& Smith (1998) have also argued for a low value of the 
transition radius in V404 Cyg.

\subsubsection{Maximum Radius for the ADAF}

At each radius $r$ in the accretion flow, there is a critical accretion
rate $\dot{m}_{\rm crit}$ above which it is not possible to have an ADAF 
solution (see Lasota 1998 for a recent discussion). 
This critical $\dot{m}_{\rm crit}$ was estimated by Abramowicz et 
al. (1995) using only bremsstrahlung cooling, by Narayan \& Yi (1995) using 
a more detailed model of cooling, and more recently by Esin (1997) and
Esin et al. (1997). We present an improved calculation here.

We are primarily interested in large radii, $r \gta 10^{3.5}$, where the 
plasma in the ADAF is essentially one-temperature (Narayan \& Yi 1995). We 
therefore set the ion and electron temperatures equal. We also assume that the 
flow is described well by the self-similar solution of Narayan \& Yi (1994),
which is well-supported by the detailed global studies of Narayan, Kato \& 
Honma (1997) and Chen, Abramowicz \& Lasota (1997). The electron density and 
temperature of the gas in the ADAF are then given by (Narayan \& Yi 1995)

\begin{eqnarray}
n_e & = & 2.00 \times 10^{19} \alpha_{\rm ADAF}^{-1} c_1^{-1} c_3^{-1/2}
m^{-1} \dot{m} r^{-3/2}~{\rm cm^{-3}},\\
T & = & 3.20 \times 10^{12} \beta c_3 r^{-1}~{\rm K}.  
\end{eqnarray}
Here $\alpha_{\rm ADAF}$ is the usual $\alpha$ viscosity parameter, 
as defined for the ADAF, and $\beta$ is the ratio of gas to total pressure, 
which we set to $0.5$ (perfect equipartition). The parameters $c_1$ 
and $c_3$ are defined to be
\begin{eqnarray}
& &c_1  =  \frac{5+2\epsilon '}{3 \alpha_{\rm ADAF}^2} g(\alpha_{\rm 
ADAF},\epsilon '),~~c_3 = \frac{2(5+2\epsilon ')}{9 \alpha_{\rm ADAF}^2} 
g(\alpha_{\rm ADAF},\epsilon '), \\
& & g(\alpha_{\rm ADAF},\epsilon ') = \left[ 1+\frac{18 \alpha_{\rm ADAF}^2}
{(5+2 \epsilon ')^2}\right]^{1/2}-1,\\
& &\epsilon ' =  \frac{1}{f_{\rm adv}} \left( \frac{5/3 - \gamma}{\gamma -1} 
\right).
\end{eqnarray}
Here $\gamma$ is the adiabatic index of the gas which, following
Esin (1997), we set to
\begin{equation}
\gamma=\frac{8-3 \beta}{6-3 \beta}=1.4444~~({\rm for}~\beta=0.5).
\label{eq:gamma}
\end{equation}
Finally, $f_{\rm adv}$ is the advection parameter of the accretion flow, 
defined as the fraction of the viscous heat energy that is advected with 
the gas.

\begin{figure}
\plotone{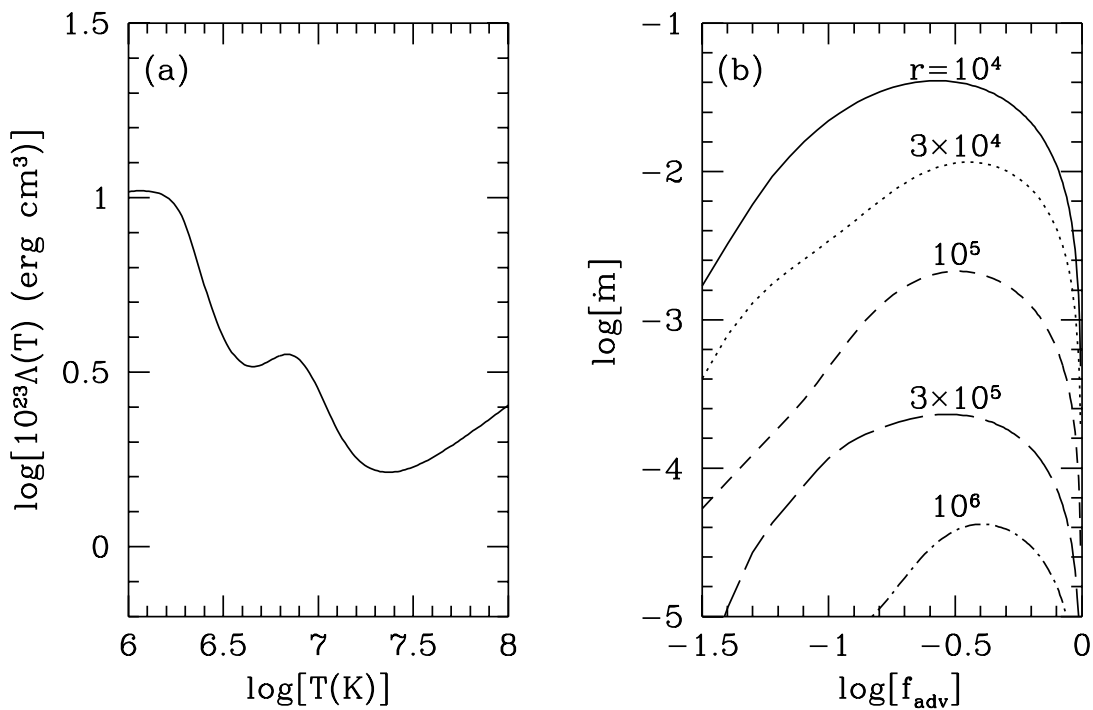}
\caption{(a) Shows the cooling function $\Lambda(T)$ used in the
calculations.  (b) Shows the mapping between the dimensionless mass
accretion rate $\dot m$ in an ADAF and the advection parameter $f_{\rm adv}$, 
at five dimensionless radii $r$.  The model assumes $\alpha_{\rm
ADAF}=0.3$ and $P_{\rm gas}=P_{\rm mag}$, or $\beta=0.5$ (i.e. exact
equipartition).  The solid, dotted, dashed, long-dashed and
dotted-dashed lines correspond to $r=10^4, 3 \times 10^4, 10^5, 3
\times 10^5$ and $10^6$, respectively.  In each case, there is a
maximum $\dot m$ up to which an ADAF is allowed, which is referred to
as $\dot{m}_{\rm crit}$.\label{fig:mdotvsf}}
\end{figure}

Given the values of $m$, $\dot{m}$, $r$, $\alpha_{\rm ADAF}$ and 
$\beta~(=0.5)$, the values of $n_e$ and $T$ depend only on  
the advection parameter $f_{\rm adv}$. 
For the moment, let us treat $f_{\rm adv}$ as a free parameter.
The self-similar solution then gives the volume heating rate due to viscous 
dissipation to be
\begin{equation}
q^+=1.84 \times 10^{21} \epsilon ' c_3^{1/2} m^{-2} \dot{m} r^{-4}~{\rm 
erg~cm^{-3}~s^{-1}},
\end{equation}
and the cooling rate to be 
\begin{equation}
q^-=n_e^2 \Lambda (T)=4.00 \times 10^{38} \epsilon ' \alpha_{\rm ADAF}^{-2} 
c_1^{-2} c_3^{-1} m^{-2} \dot{m}^2 r^{-3} \Lambda (T)~{\rm erg~cm^{-3}~s^
{-1}}, 
\end{equation}
where $n_e^2 \Lambda (T)$ is the cooling rate per unit volume of the hot 
optically thin gas. For
our calculations, we use a standard $\Lambda (T)$, shown in
Fig.~\ref{fig:mdotvsf}a, which is similar to the curve shown in
Dalgarno \& McCray (1972).

The free parameter in the above expressions is $f_{\rm adv}$.  We now
determine this parameter self-consistently.  Since $f_{\rm adv}$ is
the fraction of the viscous energy that is advected, we can equally
well say that $(1-f_{\rm adv})$ is the fraction of energy that is
radiated. (We are ignoring compressive heating, which is a reasonable
approximation at the mass accretion rates that we are considering; see
Nakamura et al. 1997, Mahadevan \& Quataert 1997.)  Thus, we determine
$f_{\rm adv}$ by solving \begin{equation} q^-=(1-f_{\rm adv})q^+.
\end{equation} Equivalently, we can solve for $\dot{m}$ as a function
of $f_{\rm adv}$: 
\begin{equation} \dot{m} = \frac{4.6 \times
10^{-18}}{\Lambda (T)}\alpha_{\rm ADAF}^2 \epsilon ' c_1^2 c_3^{3/2}
r^{-1} (1-f_{\rm adv}).  
\label{eq:mdcritadaf} 
\end{equation} 
This relation gives a mapping between $\dot{m}$ and $f_{\rm adv}$.  Note
that $\dot{m}$ is independent of $m$. Note also that if the cooling is
simply by bremsstrahlung, $\Lambda(T) \propto T^{1/2} \propto
r^{-1/2}$. We then expect $\dot{m}_{\rm crit} \propto r^{-1/2}$.  This
dependence was derived by Abramowicz et al. (1995) and Narayan \& Yi
(1995). We use a more realistic $\Lambda(T)$ and obtain a different $r$
dependence of $\dot{m}_{\rm crit}$.

Figure~\ref{fig:mdotvsf}b shows curves of $\dot{m}$ vs $f_{\rm adv}$
for $\alpha_{\rm ADAF}=0.3$, $\beta=0.5$, and various radii $r$.  At
each $r$, we see that there is a maximum $\dot{m}$ up to which
solutions for $f_{\rm adv}$ exist, but above which there is no
solution at all.  (Compare with a similar curve shown in Fig.~6a of
Esin et al. 1997). The maximum $\dot{m}$ is what we call $\dot{m}_{\rm
crit}$. For $\dot{m} < \dot{m}_{\rm crit}$, there are two solutions
for $f_{\rm adv}$. The solution with the larger $f_{\rm adv}$
(i.e. advection-dominated) is the stable ADAF branch (Narayan \& Yi
1995).  The solution with the smaller $f_{\rm adv}$
(i.e. cooling-dominated) is the SLE branch (Shapiro, Lightman \&
Eardley 1976) which is unstable (Piran 1978; Narayan \& Yi 1995).

\begin{figure}
\plotone{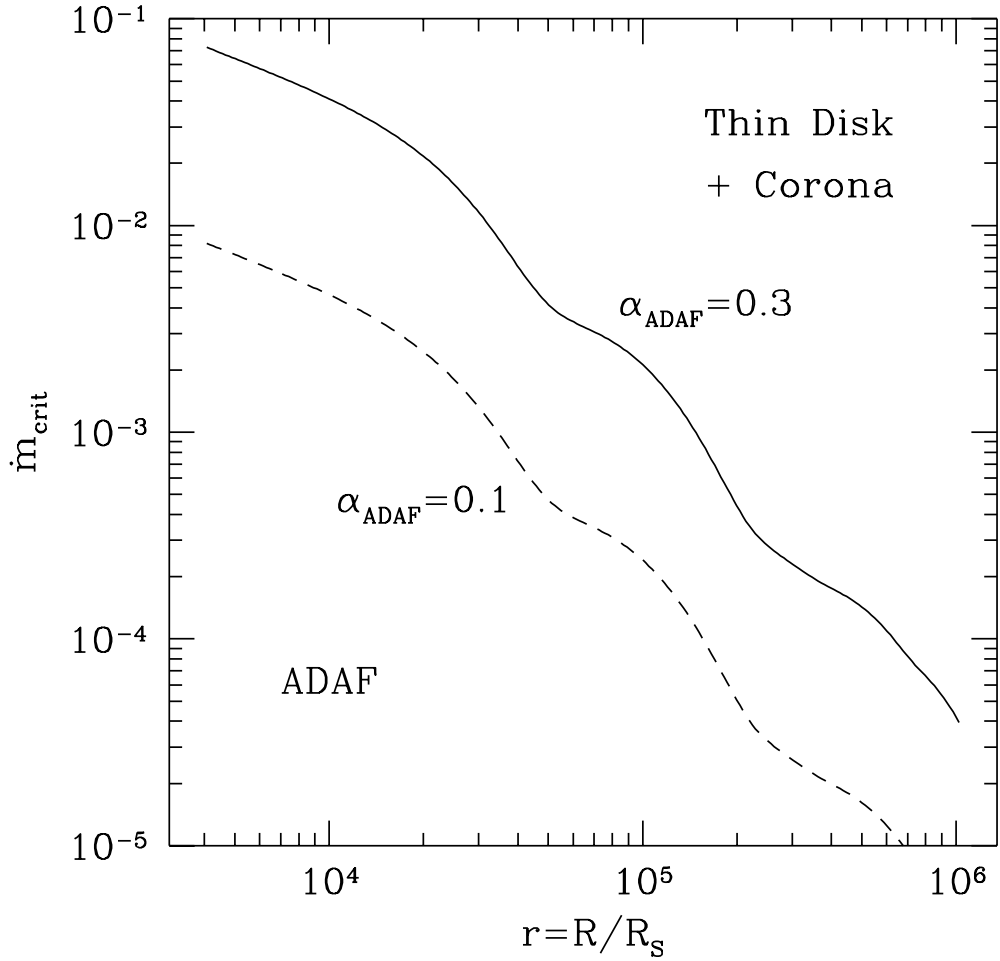}
\caption{Shows the critical mass accretion rate $\dot{m}_{\rm crit}$
as a function of radius $r$ in an ADAF for $\alpha_{\rm ADAF}=0.1$
(dashed line) and $\alpha_{\rm ADAF}=0.3$ (solid line).  An ADAF is
allowed only below and to the left of these lines.\label{fig:mdcritadaf}}
\end{figure}

Figure~\ref{fig:mdcritadaf} shows $ \dot{m}_{\rm crit}$ vs $r$ for two 
different values of $\alpha_{\rm ADAF}$. The wiggles in the curves are the
result of the various features in the cooling function $\Lambda (T)$.
Leaving the wiggles aside, we see that $\dot{m}_{\rm crit}$ decreases 
monotically with increasing $r$. 
Now, Fig.~\ref{fig:mdcritadaf} could equally well be described
as a plot of the maximum radius $r_{\rm max}$ out to which an ADAF 
is allowed for a given $\dot{m}$. 
In other words, the horizontal axis could be interpreted as showing $r_{
\rm max}$ as a function of $\dot{m}$ along the vertical axis. 
We see that $r_{\rm max}(\dot{m})$ is a monotically increasing function of
decreasing $\dot{m}$. For a given
$\dot{m}$, a pure ADAF flow is allowed only for $r < r_{\rm max}(\dot{m})$; 
beyond $r_{\rm max}(\dot{m})$ a
thin disk plus an ADAF-like corona is likely to be present
(cf Esin et al. 1997).

\subsubsection{Complete Prescription for $r_{\rm tr}$}

\begin{figure}
\plotone{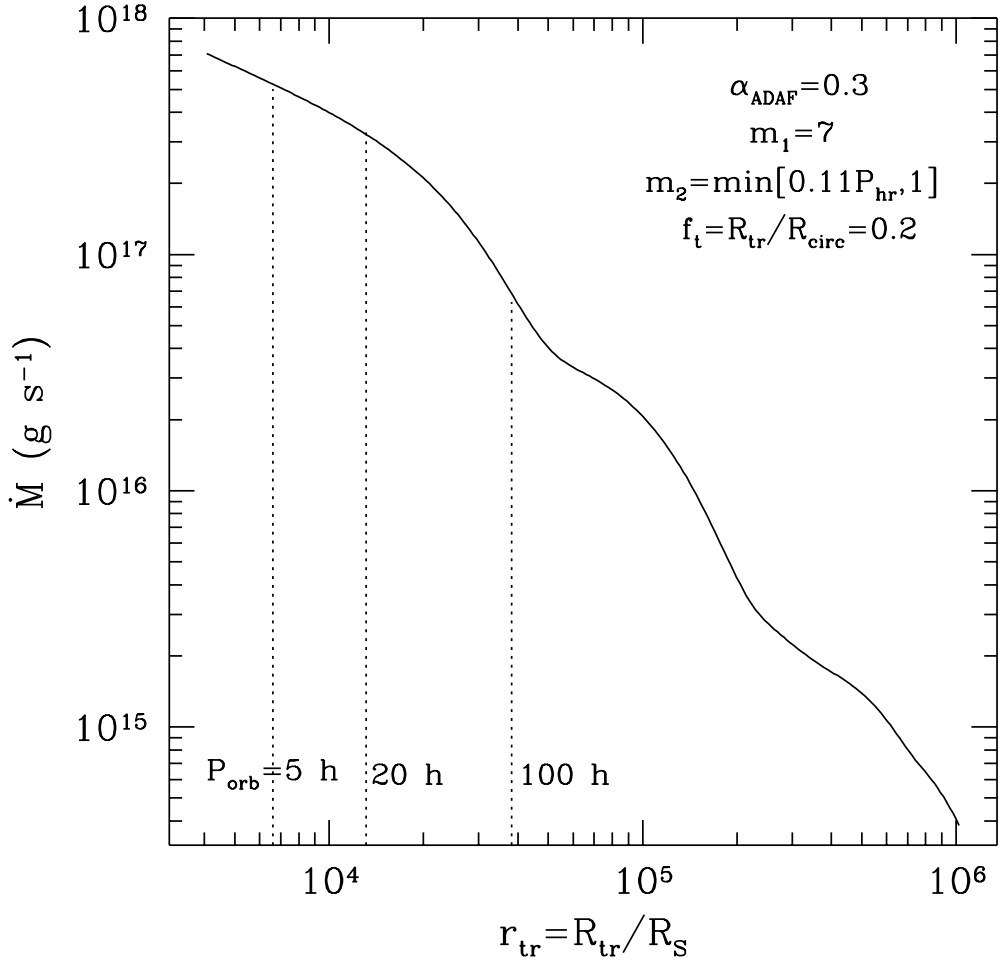}
\caption{Shows the prescription we propose for the transition radius
$r_{\rm tr}$.  The plot corresponds to a LMBHB with $\alpha_{\rm
ADAF}=0.3$, $m_1=7$, $m_2=\min[0.11 P_{\rm hr},1]$, and $f_t \equiv
R_{\rm tr}/R_{\rm circ}=0.2$. The vertical dotted lines show the
location of the transition radius, as determined from stream dynamics,
for three different orbital periods. The solid line is the maximum
accretion rate above which an ADAF is not allowed.  Equivalently, for
a given accretion rate $\dot{M}$, the solid line shows the maximum
radius $r_{\rm max}$ to which the ADAF can extend.\label{fig:critconst}}
\end{figure}

We now combine the ideas described in \S 2.1.1 and 2.1.3. 
Figure~\ref{fig:critconst} shows the results for a LMBHB with 
$\alpha_{\rm ADAF}=0.3$, $m_1=7$, $m_2=\min[0.11 P_{\rm hr},1]$ 
and the choice $f_t=0.2$.
The abscissa shows the transition radius $r_{\rm tr}$ and the ordinate shows
the mass accretion rate $\dot{M}$ in the accretion flow.
When $\dot{M}$ is low, the transition radius is
determined by stream dynamics (\S 2.1.1) and is a function only of the
orbital period of the binary. This is shown by the three vertical dotted 
lines which correspond to $P_{\rm orb}=5$ hr, $20$ hr and $100$ hr, 
respectively.

However, at each value of $P_{\rm orb}$, there is a critical $\dot{M}$ above
which an ADAF is not allowed at the value of $r_{\rm tr}$ determined 
by stream dynamics. At such high values of $\dot{M}$, the thin disk
has to extend down to a smaller radius, namely to the radius 
$r_{\rm max}$ calculated in \S 2.1.3 and shown by the solid line 
in Fig.~\ref{fig:critconst}. 
The critical $\dot{M}$ at which 
$r_{\rm tr}$ switches from the value given by stream dynamics to 
$r_{\rm max}$ differs for different orbital periods, as  
seen from Fig.~\ref{fig:critconst}.

The dotted lines in Fig.~\ref{fig:critconst} correspond to the specific
value $f_t=0.2$. Since $R_{\rm tr}$ is proportional to $f_t$ 
(Eq.~(\ref{eq:rtr})), increasing or decreasing $f_t$ would shift the 
dotted lines to the right or left, respectively.

\subsection{A New Criterion for the Global Stability of LMBHBs}

\begin{figure}
\plotone{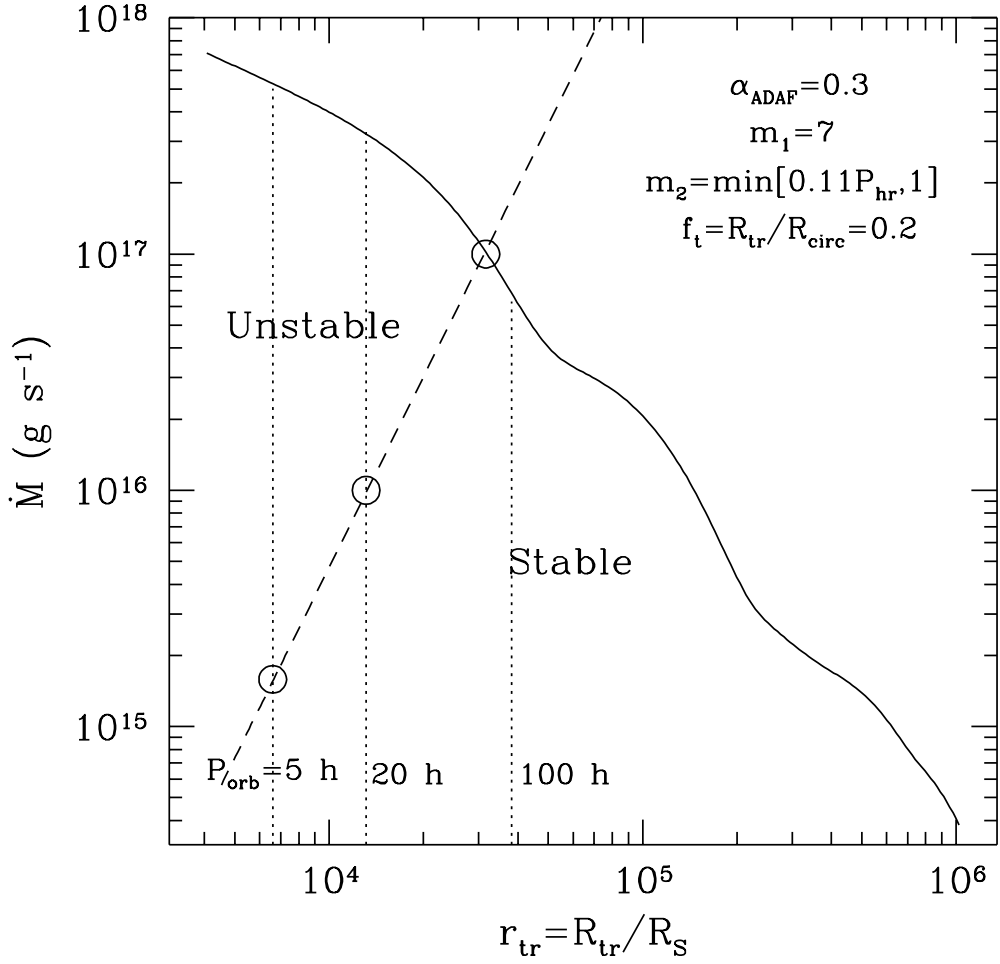}
\caption{Similar to Fig.~4  but shows in addition the limit on the
transition radius for stability of the outer disk (dashed line).  For
a given $\dot M$, the disk is stable if the transition radius is to
the right of the dashed line, and unstable if it is to the left.  The
three open circles show cases of marginal stability for three
different values of $P_{\rm orb}$ and $r_{\rm tr}$.  (See the text for
details.)\label{fig:critconst2}}
\end{figure}

According to the DIM, the accretion disk of a LMBHB is stable if the 
critical radius $r_{\rm crit}$ of the system (Eq.~(\ref{eq:recrit})) is 
smaller than the transition radius $r_{\rm tr}$. 
Figure~\ref{fig:critconst2} is similar to Fig.~\ref{fig:critconst} but 
with an additional dashed line which shows the location of $r_{\rm crit}$ as a 
function of $\dot{M}$. Any system that lies to the right of the dashed line
is stable, while any system to the left of the dashed line is unstable. 

The open circles in Fig.~\ref{fig:critconst2} show cases of marginal 
stability for three orbital periods. 
The lowermost circle shows the maximum $\dot{M}$ up to which a system 
with an orbital period of 5 hr would be globally stable. The middle circle
is the corresponding maximum $\dot{M}$ for a 20 hr binary.
The upper circle, which corresponds to $P_{\rm orb} \sim 70$ hr, 
$\dot{M}=10^{17}~{\rm g~s^{-1}}$, is a critical case. This is the limiting
$\dot{M}$ for all systems with $P_{\rm orb} \geq 70$ hr.
In all such systems, the transition radius $r_{\rm tr}$ is
determined, not by stream dynamics, but by the limiting value
$r_{\rm max}$.
At accretion rates $\dot{M} > 10^{17}~{\rm g~s^{-1}}$ (for the set
of parameters chosen in Fig.~\ref{fig:critconst2}), a LMBHB, whatever 
its period, is unstable because its transition radius is necessarily to  
the left of the dashed line.

\begin{figure}
\plotone{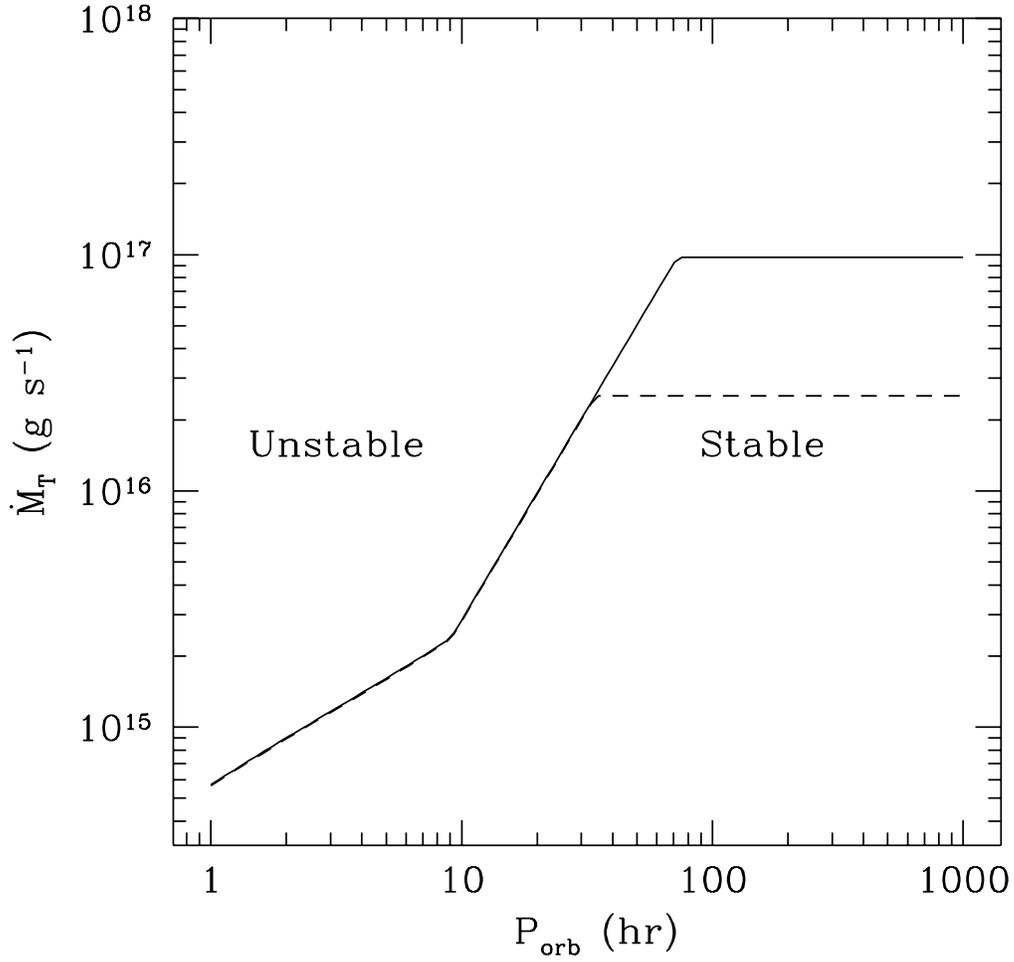}
\caption{Stability criterion (solid line) for a LMBHB with the
following parameters: $\alpha_{\rm ADAF}=0.3$, $m_1=7$, $m_2=\min[0.11
P_{\rm hr},1]$ and $f_t \equiv R_{\rm tr}/R_{\rm circ}=0.2$.  A LMBHB
of a given orbital period is unstable if it accretes mass at a rate
larger than that indicated by the solid line.  The dashed line shows
the same stability criterion if $\alpha_{\rm ADAF}=0.1$. The parameter
space for stable LMBHBs is reduced in that
case.\label{fig:critresult}}
\end{figure}

The above considerations can be summarized by the simple criterion
shown in Fig.~\ref{fig:critresult}, which corresponds to the set of 
parameters used in Figs.~\ref{fig:critconst} and \ref{fig:critconst2}. 
A LMBHB with a given orbital period is unstable if
its mass transfer rate $\dot{M}_T$ is above the solid line and stable if
$\dot{M}_T$ is below this line (for $\alpha_{\rm ADAF}=0.3$). 
If the system is stable, it will be a faint 
persistent LMBHB and very difficult to detect. 
The break in the solid line at $P_{\rm orb} \sim 70$ hr occurs because
for $P_{\rm orb} > 70$~hr
the transition radius $r_{\rm tr}$ is determined by $r_{\rm max}$ rather
than by stream dynamics.
The second break at $\sim 10$ hr corresponds to the transition from
$m_2=0.11 P_{\rm hr}$ to $m_2=1$.
Note that faint persistent systems are more likely to exist at large orbital
periods because the stability criterion is less stringent there.

\section{Comparison with Observed Transient Systems}

The purpose of this section is to establish that transient systems could
be
limited to a fairly narrow instability band in the $P_{\rm
orb}-\dot{M}$ plane.  Outside this band, the systems would be 
either hot and
steady (at high accretion rates) or cold and steady (at low accretion
rates).

We also establish that known transient systems are all found within
the instability band. This test requires a knowledge of the orbital
periods and mass accretion rates of the various BH SXTs.  The orbital
periods of the seven BH SXTs of interest\footnote{We do not consider
4U 1543-47 because there is insufficient spectral data on this newly
established BH SXT (Orosz et al. 1998).}  are well known and are
listed in Table~\ref{tab:systems}. The accretion rates in the systems
are, however, more uncertain and are necessarily model-dependent. We
describe below the method we use to estimate $\dot{M}$.

\subsection{Accretion Rates in BH SXTs}

The standard procedure to estimate the average mass accretion rate in
a SXT is based on the idea that SXTs mainly store mass during
quiescence and accrete it suddenly during an outburst (e.g. McClintock
et al. 1983, van Paradijs 1996).  The rate at which mass accumulates
in the disk, $\dot{M}_{\rm accum}$, can be estimated by measuring the
integrated flux from the system during an outburst (\S 3.1.1). This is
a reliable method since there is good evidence that SXTs accrete via a
thin disk in outburst (e.g. Tanaka \& Shibazaki 1996), so that the
radiative efficiency during outburst is well understood.  However, the
successful application of ADAF models to BH SXTs in quiescence
(Narayan, Barret \& McClintock, 1997) shows that these systems accrete
substantial amounts of mass even during quiescence via an
inefficiently radiating ADAF (\S 3.1.2).  Therefore, the total
accretion rate in a system is the sum
\begin{equation} \dot{M}_{\rm
tot}=\dot{M}_{\rm accum}+\dot{M}_{\rm ADAF},  
\label{eq:mdtot}
\end{equation} 
where $\dot{M}_{\rm ADAF}$ is the accretion rate during quiescence
(assumed to be constant over the inter-outburst time).  The total mass
accretion rate, $\dot{M}_{\rm tot}$, must be equal to the mass
transfer rate from the secondary, $\dot{M}_T$, provided all the
transferred mass is accreted by the primary.  In what follows, we
derive $\dot{M}_{\rm accum}$ and $\dot{M}_{\rm ADAF}$ for the seven BH
SXTs listed in Table~\ref{tab:systems}.

\subsubsection{Mass Accumulation Rates}

The accumulation rate $\dot{M}_{\rm accum}$
can be deduced from the total fluence, $\Delta F$, of one of
its outbursts and the inter-outburst time, $\Delta t$:
\begin{equation} 
\dot{M}_{\rm accum}=\frac{\Delta F}{\Delta t~\eta~c^2},
\label{eq:mdstor}
\end{equation}
where $c$ is the speed of light and $\eta$ is the radiative efficiency of 
accretion during outburst. We assume $\eta=0.1$, which is appropriate for a 
normal thin accretion disk around a Schwarzschild or slowly 
spinning black hole.

The observational data and the system parameters that we use to
estimate the accumulation mass accretion rates are listed in
Table~\ref{tab:systems}.  Our data are a little different (mainly
distances and outburst fluences) from those used by van Paradijs
(1996). For A0620-00 and V404 Cyg, more than one outburst has been
observed, giving $\Delta t$, so that firm estimates for $\dot{M}_{\rm
accum}$ may be derived.  For the other systems, only upper limits to
$\dot{M}_{\rm accum}$ are deduced because we have only lower limits on
$\Delta t$.  The full set of estimates for $\dot{M}_{\rm accum}$ are
listed in Table~\ref{tab:7mdot}. Our estimate for A0620-00 is in good
agreement with the estimate of McClintock et al. (1983).

\subsubsection{Accretion Rates in ADAFs}

We estimate the ADAF accretion rates in quiescence, $\dot{M}_{\rm
ADAF}$, by fitting ADAF spectral models to the observational data for
each system. 
Note that the outer thin disks do not contribute
significantly to the quiescent spectra of the systems we consider
so that the results
are insensitive to the precise choice of the transition radius.

Spectral models of A0620-00, V404 Cyg and Nova Mus 91 were constructed
by Narayan et al. (1996, 1997a), and J1655-40 was similarly modeled by
Hameury et al. (1997b). We extend the previous work on these systems
using an improved version of the ADAF model. We also construct new
ADAF models of GRO J0422+32, GS2000+25 and Nova Oph 77 in quiescence.

The new version of the ADAF model is described in Narayan et al.
(1998). One of the main improvements relative to older versions
is that the model now consistently includes adiabatic
compressive heating of electrons in the energy equation (Nakamura
et al. 1997). It also uses for the flow dynamics the general relativistic 
global solutions calculated by Popham \& Gammie (1998).

\begin{figure}
\plotone{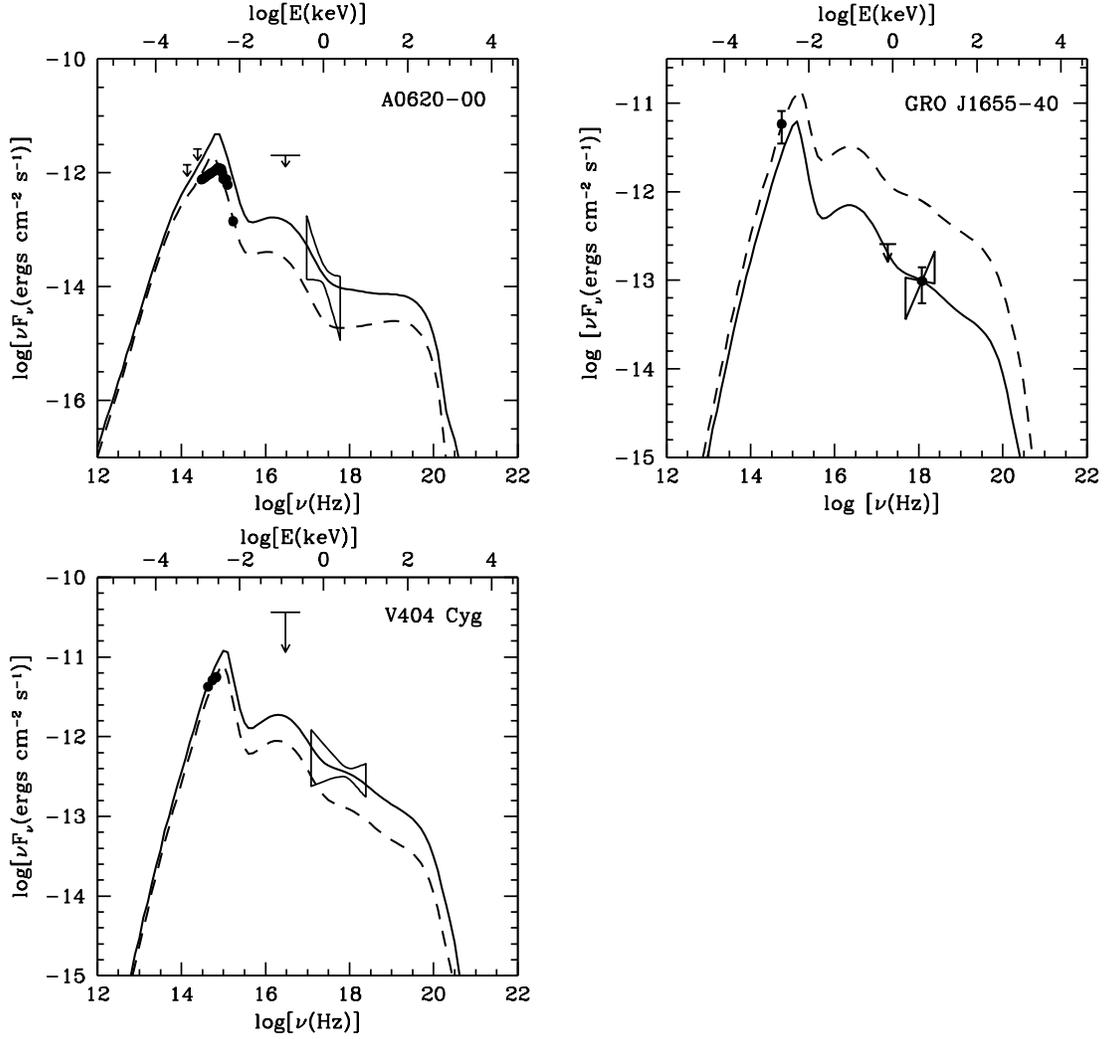}
\caption{Model spectra of A0620-00, J1655-40, and V404 Cyg.  The solid
lines show models in which the accretion rate is adjusted to fit the
X-ray data, and the dashed lines correspond to models in which the
accretion rate is adjusted to fit the optical data.\label{fig:4spec}}
\end{figure}

\begin{figure}
\plotone{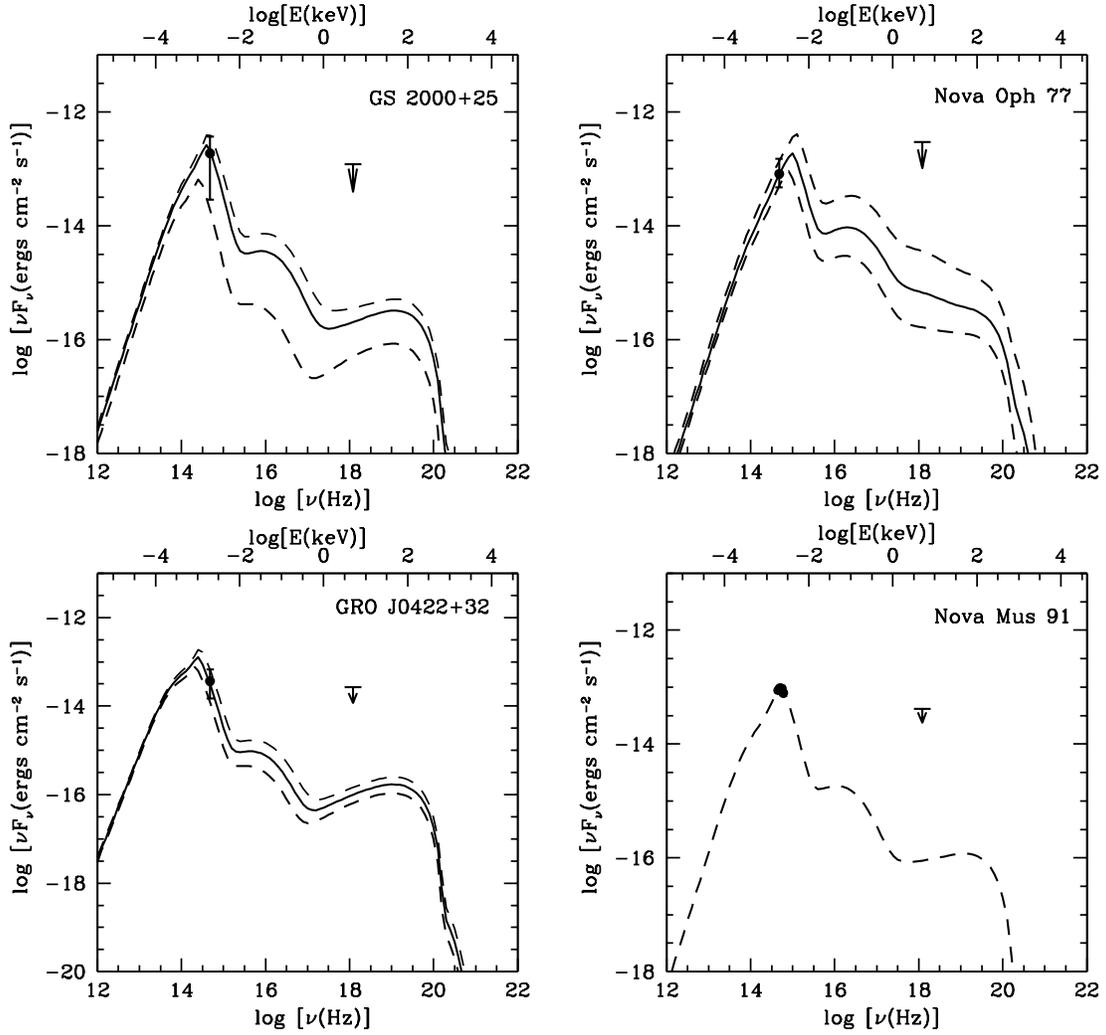}
\caption{Model spectra of GS2000+25, Nova Oph 77, GRO J0422+32, and
Nova Mus 91.  The solid and dashed lines show models with various
accretion rates which cover the range of uncertainties in the optical
fluxes.  All the models satisfy the X-ray upper limits.\label{fig:3spec}}
\end{figure}

The modeling technique used in previous studies was to fit the
quiescent X-ray flux by adjusting $\dot{M}_{\rm ADAF}$, keeping the
other model parameters fixed at the following canonical values (see
Narayan, Mahadevan \& Quataert 1998 for a discussion): $\alpha_{\rm
ADAF}=0.3$, $\beta =0.5$, $\gamma=1.4444$ (see Eq.~(\ref{eq:gamma})).
We use the same modeling technique here to estimate $\dot{M}_{\rm
ADAF}$ in those systems for which there is an X-ray flux in
quiescence. In the remaining systems, we fit the models to optical
data. The estimates of $\dot{M}_{\rm ADAF}$ are given in
Table~\ref{tab:7mdot} and the spectral fits are shown in
Figs.~\ref{fig:4spec} and ~\ref{fig:3spec}. Note that the estimates of
$\dot{M}_{\rm ADAF}$ would be reduced if $\alpha_{\rm ADAF}$ was
reduced, for instance, from $0.3$ to $0.1$.

For A0620-00, V404 Cyg, and GRO J1655-40 we used the observational
data summarized by Narayan et al. (1996, 1997a) and Hameury et
al. (1997b) and fitted the X-ray fluxes (solid lines,
Fig.~\ref{fig:4spec} ).  We also constructed models which fit the
optical data points (dashed lines, Fig~\ref{fig:4spec}) to get some
idea of the error in the estimate of $\dot{M}_{\rm ADAF}$.  The
resulting estimates are listed in Table~\ref{tab:7mdot}. Note that in
GRO J1655-40, the ``disk'' contributes only $\sim 5 \%$ of the total
optical flux (dominated by the $F$-type secondary), so that the
constraint from the optical data is rather weak (Hameury et
al. 1997b).

The remaining systems have only X-ray upper limits in quiescence and
we are limited to fitting the optical data. For Nova Mus 91 we used
the optical data summarized in Narayan et al. (1996).  For the other
three systems, namely GRO J0422+32, GS2000+25 and Nova Oph 77, we
collected optical data from the literature.  Due to the faintness of
the sources, the emission in the optical is somewhat poorly
constrained. We derived the net optical fluxes due to accretion, by
applying a correction for interstellar extinction at the wavelength of
observation (with a standard proportionality constant between the
extinction and the color excess $R_V=3.1$; Cardelli, Clayton \& Mathis
1989) and subtracting the estimated flux of the secondary.  The net
optical fluxes that we deduce in these systems are listed in
Table~\ref{tab:3obs}.  For GRO J0422+32 and GS2000+25, the uncertainty
comes mainly from the contamination of the secondary, while for Nova
Oph 77 the uncertainty is primarily from the intrinsic variability of
the source.

The ADAF spectral models for these four sources are shown in 
Fig.~\ref{fig:3spec}, for
a range of accretion rates which cover the uncertainties in the net optical 
fluxes. Note that all the models satisfy the X-ray upper limits quite
comfortably. 
We summarize in Table~\ref{tab:7mdot} the estimates for 
$\dot{M}_{\rm ADAF}$ inferred from the spectral models.

\subsubsection{Total Mass Accretion Rates}

The last column in Table~\ref{tab:7mdot} gives our estimates of 
$\dot{M}_{\rm tot}$, namely the sum of $\dot{M}_{\rm accum}$ and 
$\dot{M}_{\rm ADAF}$. An arbitrary factor of $1.5$ has been 
applied to the lower and upper bounds of the estimates
to reflect the fact that we do not trust
our estimates to an accuracy better than a factor $2$ typically.  
A look at Table~\ref{tab:7mdot} shows that in
most systems, $\dot{M}_{\rm ADAF}$ is comparable to or greater
than $\dot{M}_{\rm accum}$ (for which we have mainly upper limits), 
so that $\dot{M}_{\rm tot}$ is effectively a firm estimate. 
For Nova Mus 91 and GS2000+25, in which
$\dot{M}_{\rm accum} > \dot{M}_{\rm ADAF}$, 
only upper limits to $\dot{M}_{\rm tot}$ are deduced. Note, however, that
$\dot{M}_{\rm tot}$ cannot go below $\dot{M}_{\rm ADAF}$, which is
$\sim 2 \times 10^{15}~{\rm g~s^{-1}}$ in these two systems.

Our estimates of $\dot{M}_{\rm tot}$ are in good
agreement with the evolutionary calculations of Pylyser \& Savonije
(1988) for A0620-00 and King (1993) for V404 Cyg (closer to his maximum mass 
solution).  
Several complications may affect our result for J1655-40. The source 
remained a persistent X-ray source for
several months after its April 1996 outburst (e.g. Hynes et al. 1998)
and this has not been taken into account in the fluences listed in 
Table~\ref{tab:systems} and used to estimate $\dot{M}_{\rm accum}$ 
(Chen et al. 1997).
This could explain, at least in part, the relatively low estimate obtained for 
$\dot{M}_{\rm accum}$ (see Table~\ref{tab:7mdot}). 
The evolutionary status of J1655-40 is also very peculiar (see e.g. Kolb 1998).
In addition, Zhang, Cui \& Chen (1997) have argued that the
central black hole in J1655-40 is a Kerr black hole in fast prograde
rotation, so that the efficiency of accretion in the system is more
likely to be $\eta \sim 0.4$ rather than $0.1$. Taking this into
account would reduce $\dot{M}_{\rm accum}$ even more
in this system. Finally, the mass loss from the
secondary could be related to the jet activity in this
source (Hjellming \& Rupen 1995, 
King - private communication) which would make it difficult to get 
an independent estimate of $\dot{M}_{\rm tot}$.

\subsection{On the Stability of Known BH SXTs}

\begin{figure}
\plotone{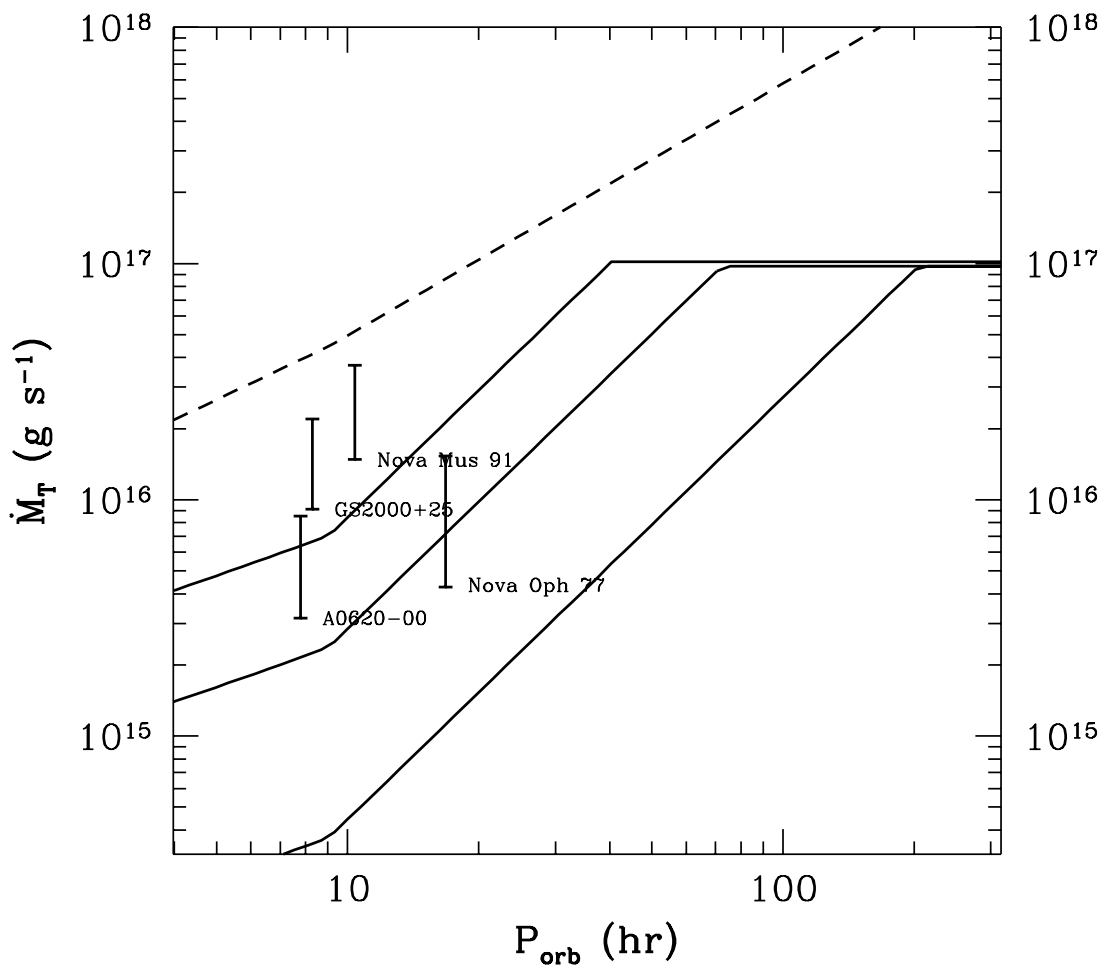}
\caption{Shows the three regions of stability for a LMBHB with
$\alpha_{\rm ADAF}=0.3$, $m_1=7$, and $m_2= \min[0.11 P_{\rm
hr},1]$.  The three solid lines correspond, from above, to
$f_t=0.3,0.2,0.1$.  A system which lies below these lines is stable
and will be a faint persistent LMBHB.  The dashed line corresponds to
the stability criterion of van Paradijs (1996).  A system that is
above this line is also stable and will be a bright persistent LMBHB.
Systems that lie in between the dashed and solid lines are unstable
and will be seen as BH SXTs.  The error bars correspond to estimates
of $\dot M_T$ of four BH SXTs.  These systems lie in the unstable
zone, as required, provided $f_t \lta 0.25-0.2$.  The remaining three
BH SXTs are not plotted as their masses are different from the choices
made here.\label{fig:critobs}}
\end{figure}

Four of the seven BH SXTs listed in 
Table~\ref{tab:systems}, namely A0620-00, GS2000+25, Nova Mus 91 and 
Nova Oph 77, have mass characteristics relatively close to those of
a `typical' system with $m_1=7$ and $m_2= \min[0.11 P_{\rm hr},1]$.
The estimated ranges of $\dot{M}_{\rm tot}$ in these four systems are shown 
in Fig.~\ref{fig:critobs} as errorbars in the $P_{\rm orb}-\dot{M}_T$ plane. 
The three solid lines are stability criteria 
as in Fig.~\ref{fig:critresult} but for three choices of the stream dynamics 
parameter, $f_t=0.3, 0.2, 0.1$ (\S 2.1.4).

Observations of $H_\alpha$ emission lines imply that
transition radii have to be relatively small ($f_t \lta 0.2-0.25$) in
quiescent BH SXTs (\S 2.1.2). 
Figure~\ref{fig:critobs} shows that for such values of
$f_t$, the four systems are indeed unstable according to 
the stability criterion. 
The estimates of $\dot{M}_{\rm tot}$ in GS2000+25 and Nova Mus 91
are subject to uncertainty and may well go down. However, they will not do so
by more than an order of magnitude (cf \S 3.1.3). 
Therefore, these systems are likely to remain in the unstable zone.
For the three other BH SXTs, namely J0422+32, J1655-40 and V404 Cyg, the
stability criteria have to be computed for their specific mass 
characteristics. We find that J0422+32 and J1655-40 
are unstable, so long as $f_t \lta 0.25$, although
the evidence is less strong than in the previous four cases 
(the estimates of $\dot{M}_{\rm tot}$ are not firmly above the stability 
line corresponding to $f_t=0.2$). 
This could be explained in J0422+32 by the fact
that the mass of the central black hole,
which has a substantial effect on the stability criterion, is not well 
known (Beekman et al. 1997; we take $m_1=12$). 
In the case of J1655-40, we expect 
$\dot{M}_{\rm tot}$ to be larger than our estimate (because some 
activity was not taken into account, cf \S 3.1.3), 
which could easily drive the system firmly into the region of instability.
Our analysis shows that V404 Cyg would be stable if $f_t=0.2$ and 
unstable if $f_t \lta 0.1$. This is consistent 
with the maximum value for $f_t$ deduced from observations of this 
system (\S 2.1.2).

These results strongly suggest that the truncated disks found in BH
SXTs are unstable and that the systems will undergo the
thermal-viscous instability.  All of these systems are indeed
transients and do show large variability. Thus the same
mechanism could drive outbursts of both DNe and SXTs
(van Paradijs \& Verbunt 1984). It is not possible, however, to firmly
rule out the possibility that at least some of the disks are stable
since the systems generally lie close to the stability region.
In this case SXT outbursts could be triggered by variations
in the mass-transfer rate (but outbursts would still be due to
a thermal-viscous instability) as proposed by \cite{lhh95}.

A criterion for the stability of (hot) irradiated disks in SXTs was
derived by van Paradijs (1996) (see however Dubus et al. 1998).  The
criterion is shown as a dashed line in Fig.~\ref{fig:critobs}. All
systems above this line are globally stable according to this
criterion. Van Paradijs did not include the additional accretion
occurring via the ADAF in quiescent BH SXTs.  One might worry that
including it may modify his results.  However, as
Fig.~\ref{fig:critobs} shows, the accretion rate estimates
($\dot{M}_{\rm tot}$) for A0620-00, GS2000+25, Nova Mus 91 and Nova
Oph 77, which take into account the additional accretion via the ADAF,
are all below van Paradijs' stability line. The same is true for
J0422+32, J1655-40 and V404 Cyg (not shown).

To summarize, the $P_{\rm orb}-\dot{M}_T$ plane is divided into three
zones.  The region to the lower right in Fig.~\ref{fig:critobs}, below
the solid line (which line we choose depends on the choice of $f_t$),
corresponds to faint persistent systems in which the disk is in a
globally stable cool state.  No such systems have been found yet,
but they may well exist and remain undetected, as they are likely
to be very dim. The middle zone, between the solid and dashed lines
corresponds to an unstable disk. Such systems are expected to be
transients, and indeed all known dynamically confirmed BH SXTs are found
in this region. Finally, the region above the dashed line corresponds
to luminous persistent systems in which the irradiated disk is in a
globally stable hot state.  The persistent black hole binaries, Cyg X-1
and LMC X-3, both lie in this region. Neither is a LMBHB, but van
Paradijs' criterion applies equally to high mass binaries (provided they
accrete from a disk). A number of other persistent black hole binaries
may also lie in this zone, but there is not yet enough dynamical
information to determine their status.

Figure~\ref{fig:critobs} shows that if transition radii in quiescent BH
SXTs were slightly larger (i.e if $f_t$ were larger), some of the
systems would become globally stable and would correspond to faint persistent
LMBHBs.  Do such systems exist in the Galaxy?

\section{A Galactic Population of Dim Accreting Binary BHs ?}

To answer the above question,
we use  models of binary evolution to predict the accretion rates in binary 
systems as a function of their orbital periods and check if
any of the evolutionary tracks cross the lowermost zone in 
Fig.~\ref{fig:critobs}.

\subsection{Mass Transfer from Binary Evolution Models}

The standard theory of binary evolution relies on the Roche-lobe model
for the description of mass transfer in binary star systems.  Depending
on which mechanism drives the mass transfer, whether it is loss
of orbital angular momentum through gravitational radiation and
magnetic braking, or expansion of the donor as it evolves away from the
main-sequence, the systems are classified as j-driven or
n-driven systems. There is a `bifurcation' orbital period
separating the two classes such that above this period we find
n-driven systems whose $P_{\rm orb}$ increase with time,
and below this period we find j-driven systems with $P_{\rm orb}$ decreasing
with time. The precise value of the bifurcation period depends on the
history of the system. 
Estimates range from $0.5-2$ days (e.g. Pylyser \& Savonije 1988, King, Kolb \& Burderi 
1996).

\begin{figure}
\plottwo{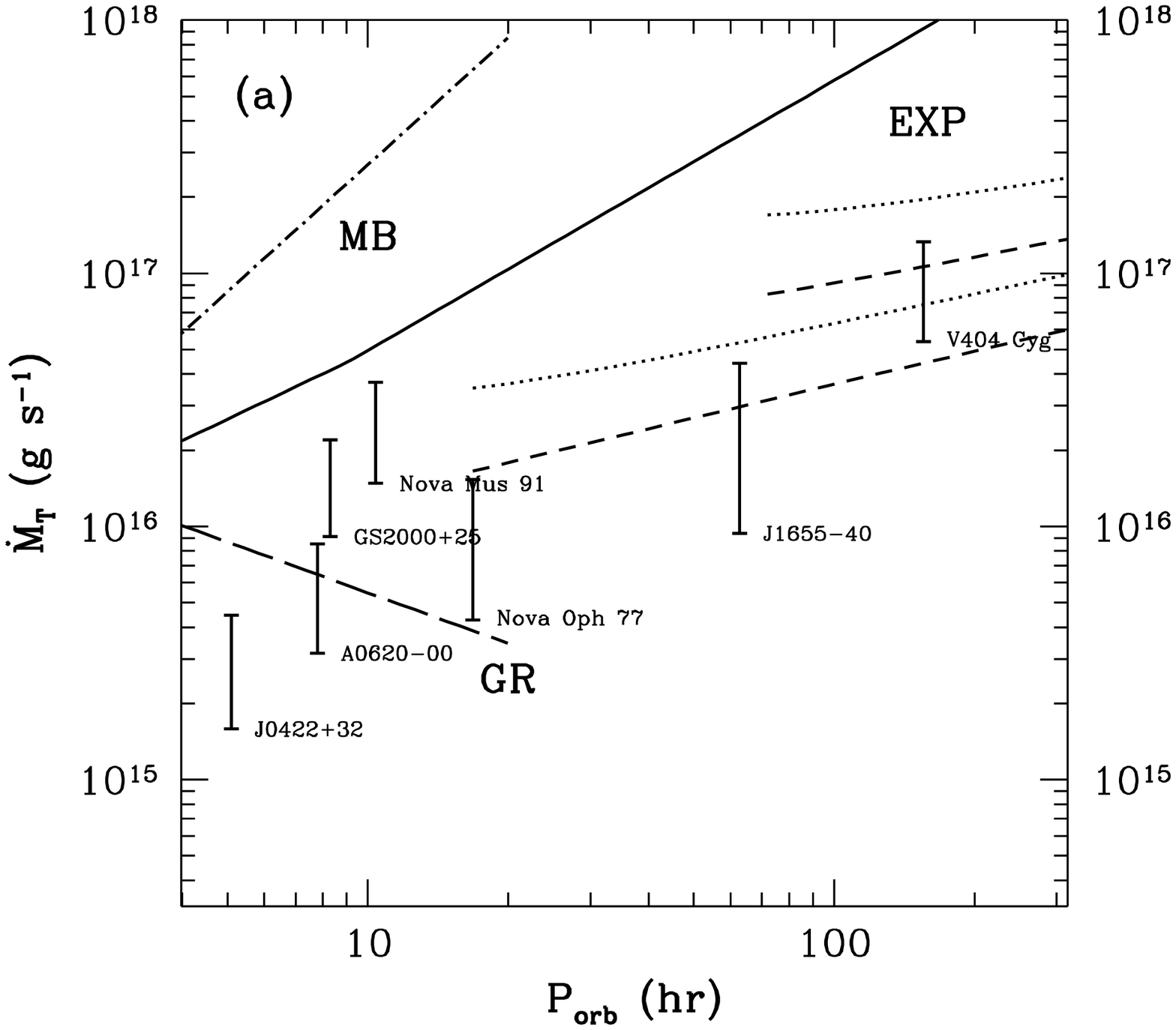}{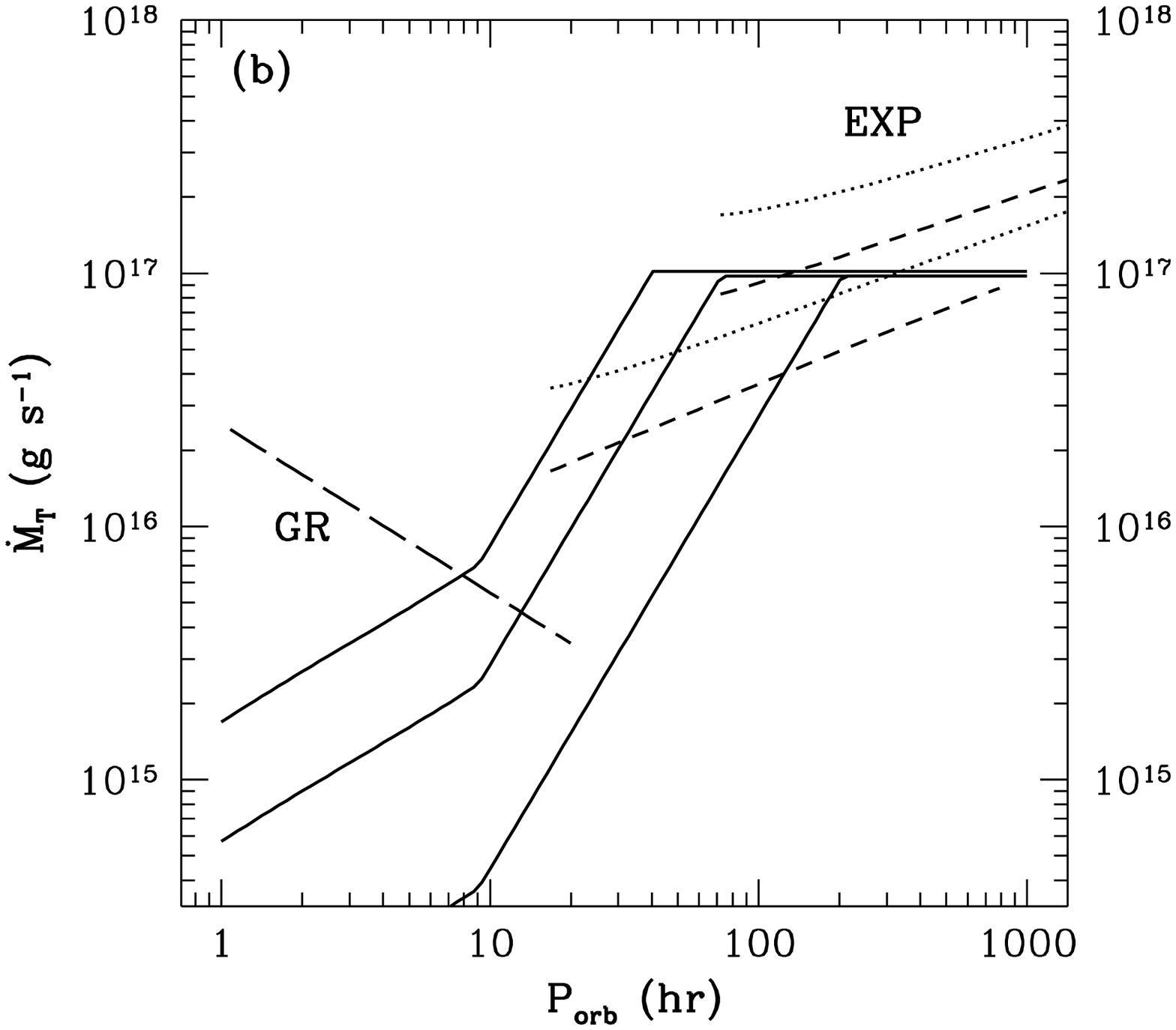}
\caption{(a) A comparison of the estimates of $\dot{M}_{\rm tot}$
(errobars) for the seven BH SXTs listed in Table~1 with the mass
transfer rates predicted by binary evolution models. At low orbital
periods, $P_{\rm orb}<20$ hr, predictions for gravitational radiation
driven mass transfer (GR, long dashed line) and magnetic braking
driven mass transfer (MB, dotted-dashed line) are shown.  At longer
orbital periods, evolutionary tracks corresponding to mass transfer
driven by secondary expansion (EXP) are shown.  Four tracks,
corresponding to secondary initial masses of $m_2=1$ (dashed lines)
and $1.5$ (dotted lines), and two different initial orbital periods
($P_{\rm orb}=17$ hr and $72$ hr) are shown. The solid line corresponds to
the stability criterion of van Paradijs (1996).  
(b) Same as (a), except that the
mass transfer rates predicted by binary evolution models (GR and EXP)
are compared to the stability criteria shown in Fig.~9.  The region
below the solid lines is stable.  Note that some segments of the
evolutionary tracks lie in the stable zone.  Systems that correspond
to these stable segments would appear as faint persistent LMBHBs.
\label{fig:critmdot}}
\end{figure}

Following King et al. (1996), we take the mass transfer rates in
j-driven systems with gravitational radiation (GR) and  
magnetic braking (MB), and n-driven system with secondary expansion (ex), to
be given by:  
\begin{equation}
\dot M_{\rm GR} = 1.27 \times 10^{14} m_1 m_2^2 (m_1 +  m_2)^{-1/3} 
                  P_{\rm d}^{-8/3} \ \ {\rm g \ s^{-1}},
\label{gr}
\end{equation}
\begin{equation}
\dot M_{\rm MB} = 3.17 \times 10^{17} m_1^{-1} m_2^{7/3} (m_1 +  m_2)^{1/3}
                  P_{\rm d}^{-2/3} \ \ {\rm g \ s^{-1}},
\label{mb}
\end{equation}
\begin{equation}
\dot M_{\rm ex} = 2.54 \times 10^{16} m_2^{1.47} P_{\rm d}^{0.93} 
                  \ \ {\rm g \ s^{-1}}.
\label{ev}
\end{equation}
Note that Eq.~(\ref{ev}) is a large oversimplification; a
substantial scatter around this prediction is expected, depending on
the precise history of the system (see Fig.~\ref{fig:critmdot} for more
accurate evolutionary tracks). 

The exact form of the magnetic braking law is not well established. 
Eq.~(\ref{mb}) corresponds to the formula due to Verbunt \& Zwaan (1981).
Other models of
magnetic braking have been proposed in the literature (see e.g. Warner
1995) but they give similar results (not differing by more than a factor of
a few).
Eq.~(\ref{mb}) does not apply, however, to systems with orbital
periods $\gta 10$ hr.  The magnetic braking model represents well
the {\sl secular} mean accretion rate of cataclysmic variables and
allows an understanding of the so called `period-gap' in the orbital
period distribution of cataclysmic binaries. (Contrary to some recent
and less recent assertions, the period gap does exist,
e.g. Hellier \& Naylor 1998.) One observes, however, a rather large
scatter of mass transfer rates deduced from observations, a large
part of which is clearly intrinsic (Warner 1995). This means that some
systems and some classes of systems transfer mass at rates
different from the secular one. These fluctuations are
compatible with the secular evolution as long as the durations of the
fluctuations are much
shorter than the characteristic evolutionary time of the system 
($\sim 10^8$ years in the case of magnetic braking).

We plot in Fig~\ref{fig:critmdot}a the predictions of the
binary evolution model for GR driven (long-dashed line, Eq.~(\ref{gr})) 
and MB driven (dotted-dashed line, Eq.~(\ref{mb})) mass transfer rates, for
orbital periods up to $P_{\rm orb}=20$ h. 
At longer periods, we plot the predictions for secondary expansion 
corresponding to detailed evolutionary tracks rather than the 
crude Eq~(\ref{ev}). The evolutionary tracks were computed 
with a semi-analytical model, described by Verbunt \& van den Heuvel (1995). 
It gives results which compare well with those of more detailed stellar 
evolution models since the properties of a (sub)giant star depend mainly on 
the mass of its core (Webbink, Rappaport \& Savonije 1983).
Four examples of evolutionary tracks are shown in Fig~\ref{fig:critmdot}a for 
binary systems with an initial mass of the secondary $m_2=1$ (dashed line) 
and $1.5$ (dotted line), with two different initial periods 
($17$ hr and $72$ hr). The mass of the primary is $m_1=7$ in each
case, but the predictions for the mass transfer rates  
increase by only $\sim 20 \%$ if the mass of the primary 
is $m_1=15$. The composition is assumed to be solar.
Note that the specific evolutionary state 
of J1655-40 is probably not accurately represented by any of the evolutionary 
tracks shown (e.g. Kolb 1998).

Also plotted in Fig~\ref{fig:critmdot}a are our estimates of
$\dot{M}_{\rm tot}$ for the seven known BH SXTs.  While the
predictions for GR driven and secondary expansion driven mass transfer
(Eq.~(\ref{gr})) appear in reasonable agreement with the estimates of
$\dot{M}_{\rm tot}$, there is clearly a problem in the case of MB
driven mass transfer (Eq.~(\ref{mb})), for which the theoretical
predictios are well above the estimates of $\dot{M}_{\rm tot}$ in
J0422+32, A0620-00, GS2000+25, Nova Mus 91 and Nova Oph 77.  One
should keep in mind that the estimates of $\dot{M}_{\rm tot}$ in
GS2000+25 and Nova Mus 91 may well go down by as much as an order of
magnitude (\S 3.1.3), which would make the discrepancy more
serious. The disagreement would be even more acute if we did not
invoke a low radiative efficiency ADAF in quiescence.

Quite apart from the inconsistency between the estimates of
$\dot{M}_{\rm tot}$ and the magnetic braking prediction, we note the
serious problem that the MB line in Fig~\ref{fig:critmdot}a lies
entirely above van Paradijs' stability criterion, shown by the solid
line .  If Eq.~(\ref{mb}) is correct, then there should be no transient
LMBHBs with $P_{\rm orb} < 20$ hr.  But five of the seven transient
systems we study have periods in this range. Equation~(\ref{mb}) is
clearly inapplicable to these systems. 

How do we resolve this discrepancy? 
There is no theoretical reason or observational evidence to
suggest that magnetic braking does not operate in LMBHBs.
There are reasons, however, to think that the companions in LMBHBs could 
be evolved (King et al. 1996). In this case Eq.~(\ref{mb}) 
becomes:
\begin{equation}
\dot M_{\rm MBev} = 3.17 \times 10^{17} m_1^{-1} \hat m_2^{7/3} 
(m_1 +  m_2)^{1/3}
                  P_{\rm d}^{-2/3} \ \ {\rm g \ s^{-1}},
\label{mbe}
\end{equation}
where $\hat m_2=m_2/m_{2,{\rm main-sequence}} \approx 0.5 - 0.6$. 
A similar formula can be deduced for GR driven mass transfer with 
an evolved donor. 
The accretion rates predicted by ``evolved'' GR and MB laws are 
typically a factor $3$ to $5$ smaller than the predictions of
Eq.~(\ref{gr}) and Eq.~(\ref{mb}). As suggested by King et al. (1996),
this would bring the predictions closer to the observational estimates.
The law of Verbunt \& Zwaan (1981) predicts relatively high accretion rates. 
Using a different law, for instance the Rappaport, Joss \& Verbunt (1983) law
(see e.g. Kalogera, Kolb \& King 1998), would bring the predictions even 
closer to the observational estimates.

There is an additional argument, based on the recurrence times of
BH SXTs, which shows that Eq.~(\ref{mb}) is inappropriate to 
predict mass transfer rates in LMBHBs. This is discussed in the next 
subsection. 

\subsection{Recurrence Times}

According to the DIM, the maximum surface density of a cold (quiescent) disk 
is (Hameury et al. 1998, see also e.g. Ludwig et al. 1994, Smak 1993):
\begin{equation}
\Sigma_{\rm max} = 13.1 ~ \alpha^{-0.85} {m_1}^{-0.37} R_{10}^{1.11}
~\rm g~cm^{-2}.
\end{equation}
The corresponding maximum quiescent disk mass is given by:
\begin{equation}
M_{\rm max} = 3.9 \times 10^{21} \alpha^{-0.85} m_1^{-0.37} 
R_{\rm D, 10}^{2.11} 
\ {\rm g},
\label{diskmass}
\end{equation}
where $R_{\rm D,10}$ is the disk outer radius in units of $10^{10}$ cm. 
As suggested by observations
(e.g. Harrop-Allin \& Warner 1996) we will assume that $R_{\rm D}$
is about 70\% of the Roche-lobe equivalent radius. 
Using an approximate formula by Paczy\'nski (1971),
\begin{equation}
R_{\rm L} \approx \left[ 0.38 - 0.2 \log \left({m_2 \over m_1} \right)\right] 
a, \ \ {\rm for} \ \ 0.3 < {m_1 \over m_2},
\end{equation}
where $a$ is the binary separation, we obtain
\begin{equation}
R_{\rm D, 10}\sim 15  m_1^{1/3} P_{\rm orb}^{2/3},
\label{drad}
\end{equation}
with the orbital period $P_{\rm orb}$ measured in days.  

There is an important difference between WZ Sge-type systems and SXTs.
During a (super)outburst of the short period dwarf nova WZ Sge, more
than $10^{24}$ g of matter was accreted onto the central white dwarf
(Smak 1993). Taking the disk radius in this system to be $\sim
10^{10}$ cm, we see from Eq.~(\ref{diskmass}) that $\alpha$ in
quiescence must be very low ($\lta 10^{-4}$), unless, due to
irradiation of the secondary, mass transfer is increased during the
outburst (Hameury, Lasota \& Hur\'e 1997). The amount of mass accreted
during SXT outbursts is also $\sim 10^{24}$~g but, since their orbital
periods are between 0.2 and 6 days, the disks are large enough to
contain the required mass even for a normal value of $\alpha$ (i.e.
$\sim 10^{-2}$).  Indeed, the secondary in WZ Sge showed evidence for
irradiation during an outburst (Smak 1993) whereas there is no
evidence for significant irradiation of the secondary in any SXT
(McClintock, private communication).

The recurrence time between two successive outbursts can be estimated
very crudely as the time required to fill up the
disk to its maximum mass:
\begin{equation}
t_{\rm rec} \lta {M_{\rm max} \over \dot M_{\rm T}
\left(1 - {\dot M_{\rm acc}
\over \dot M_{\rm T}}\right)
}
\label{rectime}
\end{equation}
where $\dot M_{\rm acc}$ and $\dot M_{\rm T}$ are the accretion rate in
quiescence at the inner disk radius (in the case of an ADAF + thin disk
accretion flow, the `inner' radius is the transition radius)
and the mass transfer rate from the secondary, respectively.  In the
DIM framework, if the quiescent disk extends down to the last stable
orbit (or to the surface of the compact body in the case of 
white dwarfs) $\dot M_{\rm
T} \gg \dot M_{\rm acc}$ (see Eq.~(\ref{eq:mdcrit-})) so that the
recurrence time is simply given by $t_{\rm rec} \lta M_{\rm max}/ \dot
M_{\rm T}$. In the presence of an inner ADAF in quiescent SXT accretion
flows (and in a similar model for WZ Sge by Hameury et al. 1997a), the
`leak' at the transition radius may require the use of
Eq.~(\ref{rectime}).
However, the term 
$\left(1 - {\dot M_{\rm acc}/ \dot M_{\rm T}}\right)^{-1}$ is generally
very close to unity because
\begin{equation}
{\dot M_{\rm acc}\left(0.2 \times R_{\rm circ}\right) \over 
\dot M_{\rm T}} \approx
0.04 m^{-1} m_2^{-7/3} m_1^{0.11} P_{\rm d}^{2.45} \alpha^{-0.04},
\end{equation}
where we use the critical accretion rate in Eq.~(\ref{eq:mdcrit-}) to
estimate $\dot M_{\rm acc}$.

In order to estimate recurrence times in the ADAF + thin disk model,
we use the mass-transfer rates given in Eqs.~(\ref{gr})-(\ref{ev}).
Combining these with Eqs.~(\ref{diskmass}) and (\ref{drad}), we find
\begin{eqnarray}
{\rm GR:} \ \ t_{\rm rec}& = & 295 \ {\rm y} \ \alpha^{-0.85} m_1^{-0.67} m_2^{-2}
                           (m_1 +  m_2)^{1/3} P_{\rm d}^{4.08}
                          \left(1 - {\dot M_{\rm acc}\over \dot M_{\rm T}}\right)
                           ^{-1},
\\
{\rm MB:} \ \ t_{\rm rec}& = & 0.3 \ {\rm y} \ \alpha^{-0.85} m_1^{1.33} m_2^{-7/3}
                           (m_1 +  m_2)^{-1/3} P_{\rm d}^{2.08}
                          \left(1 - {\dot M_{\rm acc}\over \dot M_{\rm T}}\right)
                          ^{-1},
\\
{\rm ex:} \ \ t_{\rm rec}& = & 1.47 \ {\rm y} \ \alpha^{-0.85} m_1^{0.33} m_2^{-1.47}
                           P_{\rm d}^{0.48}
                          \left(1 - {\dot M_{\rm acc}\over \dot M_{\rm T}}\right)
                           ^{-1}.
\label{trec}
\end{eqnarray}
Using mass-period relations for main sequence stars ($m_2 \approx 
0.11 P_{\rm h}$) and for subgiants ($m_2 \propto P^{-0.33}$, e.g. King 1988) 
we obtain acceptable estimates of the recurrence time
for the cases of gravitational radiation and secondary evolution. 
The recurrence time for the magnetic braking case is, however, far too short.
This has also been noticed by Romani (1998). The use of Eq.~(\ref{mbe}), 
instead of Eq.~(\ref{mb}), makes the estimated recurrence
time longer if we take $\hat m_2 \approx 0.5 - 0.6$.

There is a clear need for better predictions of the mass transfer rates 
via magnetic braking in low orbital period systems. 
Direct observational constraints on 
magnetic braking in cataclysmic variables are scarce 
(Kolb, private communication) and the extrapolation of these data 
to LMBHBs is hazardous. For these reasons, we choose to ignore the contribution
of magnetic braking to the mass transfer rate in what follows.

\subsection{Do Faint Persistent LMBHBs exist?}

Figure~\ref{fig:critmdot}b shows the same predicted mass transfer
rates as in Fig.~\ref{fig:critmdot}a, for GR driven and secondary
expansion driven mass transfer, but omitting the problematic MB case.
Three stability criteria (solid lines, $f_t=0.3, 0.2, 0.1$ from above)
are also shown, for a `typical' system with $m_1=7$ and $m_2= \min
[0.11 P_{\rm hr},1]$.  Recall that a system that is below one
of these lines will be faint and persistent.

At short orbital periods ($P_{\rm orb} \lta 20$ h), binary evolution
models predict accretion rates which are substantially above the limit
for stability.  We do not expect faint persistent LMBHBs in this range
of periods for any reasonable value of $f_t$.  This conclusion is
independent of the uncertainty on the magnetic braking law because
magnetic braking only predicts larger accretion rates.

At long periods, however, some of the evolutionary tracks do predict
accretion rates low enough to allow the existence of faint LMBHBs.
This is especially true for $f_t \sim 0.2-0.3$, but some persistent
systems are expected even if $f_t \sim 0.1$. Thus, faint persistent
LMBHBs are viable at long orbital periods, provided that transition
radii in these systems are not too small (i.e $f_t$ not $\ll 0.1$).
What would such systems look like?

\subsection{Observational Signatures of Faint Persistent LMBHBs}

The main difference between a faint persistent LMBHB and a quiescent
SXT is that $r_{\rm tr}>r_{\rm crit}$ in the former and $r_{\rm
tr}<r_{\rm crit}$ in the latter. Of course, since $r_{\rm crit}$
is a function of $\dot m$, variations of the mass-transfer rate
can move systems from one class to another.

In both kinds of system, $r_{\rm tr}$ is sufficiently large that the
emission from the outer thin disk is weak and unimportant relative to
the emission from the secondary star.  Further, the emission from the
ADAF is dominated by the regions of the flow close to the black hole
($r<30$), which means that the precise value of $r_{\rm tr}$ is
unimportant.  For both reasons, we expect the spectrum of a faint
persistent LMBHB to be very similar to that of a quiescent SXT with
the same mass accretion rate (or similar orbital period).

\begin{figure}
\plotone{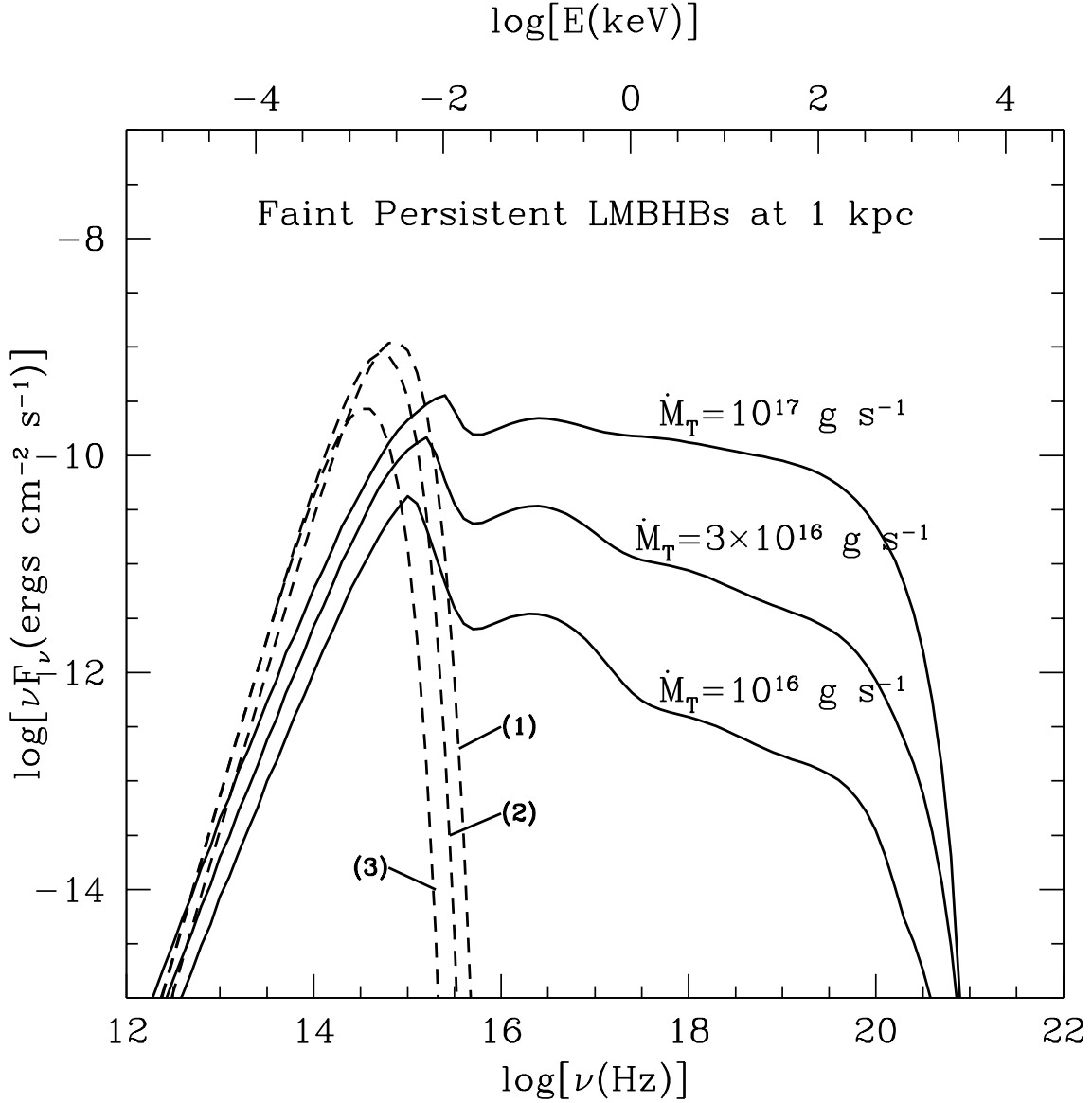}
\caption{Model spectra of faint persistent LMBHBs corresponding to
three mass accretion rates (solid lines).  The models assume that the
sources are at a distance of $1$ kpc and take $\alpha_{ADAF}=0.3$, $\beta=0.5$
and $m_1=7$.  The dashed lines show spectra of the secondary stars in the
long-period SXTs, 4U 1543-47 (1), GRO J1655-40 (2) and V404 Cyg (3). 
Faint persistent LMBHBs are expected to have similar secondaries. Note that
the optical emission from the accretion flow 
is likely to be swamped by the emission from the secondaries.
\label{fig:specfaint}}
\end{figure}

By the arguments given earlier, we expect faint persistent LMBHBs to
have relatively long orbital periods ($P_{\rm orb}>30$ hr) and fairly
large mass transfer rates ($10^{17}~{\rm g\,s^{-1}}>\dot M_T>
10^{16}~{\rm g\,s^{-1}}$).  We show in Fig~\ref{fig:specfaint}
predicted spectra of representative faint persistent LMBHBs (solid
lines) as they would appear if the sources were located at a distance
of $1$ kpc. Three different accretion rates are shown, corresponding to
a black hole mass $m_1=7$ and $\alpha_{\rm ADAF}=0.3$.  A comparison
between Fig.~\ref{fig:specfaint} and Figs.~\ref{fig:4spec} and
\ref{fig:3spec} reveals the strong spectral similarity between faint
persistent LMBHBs and quiescent SXTs.

In the three known BH SXTs with long orbital periods, namely V404 Cyg
($P_{\rm orb}=155.4$ hr), J1655-40 (62.7 hr) and 4U 1543-47 (26.9 hr),
the secondary dominates the optical and IR emission in quiescence
(this is true even in the case of the cool companion in V404 Cyg where 
the disk contributes less than about 15\% in optical and IR; Casares et al.
1993; Shahbaz et al. 1996).  We expect faint persistent LMBHBs
to have subgiant secondaries similar to those found in these three BH
SXTs. We show in Fig~\ref{fig:specfaint} model black body spectra
corresponding to the secondaries in V404 Cyg, GRO J1655-40 and 4U
1543-47.  As expected, the secondaries dominate the emission in the
optical band; the accretion flow contributes very little.  Clearly, it
would be very difficult to discover faint persistent LMBHBs via any
kind of broadband survey in the optical.  One could in principle look
for excess emission in blue and ultraviolet radiation, arising from
synchrotron radiation in the ADAF, but this is not likely to be a very
effective technique.

Of course, if any particular system is suspected to be a faint
persistent LMBHB, it would be relatively easy to verify its LMBHB
nature through observations.  The orbital motion of the secondary
could be confirmed through measurements of the Doppler shifts of its
spectral lines.  This would reveal the mass function of the binary and
may even lead to a determination of the mass of the black hole (as in
the best BH SXTs).  Also, the detection of a broad
double-humped $H_{\alpha}$ line would reveal the presence of the outer
thin accretion disk, just as in quiescent BH SXTs.  Finally,
observations in X-rays would reveal the presence of the ADAF.

Another severe problem is that even if one did discover a candidate
system, it will not be easy to confirm that it is a faint persistent
LMBHB rather than a quiescent SXT.  SXTs have typical recurrence times
of 20 to 50 years.  One would have to observe the candidate system for
at least 50 years before one could be reasonably confident that it is
not a SXT.

\section{Discussion}

Whether or not faint persistent LMBHBs exist in the Galaxy depends on
the mass transfer rates $\dot{M}_T$ and transition radii $R_{\rm tr}$
in black hole binaries.  While our predictions of the mass transfer
rates, especially via secondary expansion, are relatively robust, this
is not the case for our estimates of $R_{\rm tr}$.  We present a
specific prescription for estimating the transition radius, based on
stream dynamics and other considerations.  Observations of
$H_{\alpha}$ emission lines provide only upper limits to the values of
$R_{\rm tr}$.  In the absence of a more complete theory of gas
evaporation from the thin disk to the ADAF, the location of the
transition radius remains a major uncertainty of the model.  Indeed,
until this issue is resolved it is difficult to estimate what fraction
of the LMBHB population in the Galaxy might correspond to faint
persistent systems.

Another uncertainty is related to time variability of the mass
transfer rate.  We have shown that the known SXTs have outer disks
that are unstable to the thermal-viscous instability of the DIM, and
therefore that these systems would be unstable even with a steady mass
transfer from the secondary.  However, Kuulkers et al. (1996) recently
emphasized the striking similarities between the lightcurves of WZ Sge
type dwarf novae and SXTs.  In the case of the WZ Sge type systems,
one needs to invoke a variable mass transfer rate, with an enhancement
during the outburst, if one wishes to avoid using extremely low values
of the viscosity parameter $\alpha$ (see Hameury et al. 1997a).  If
LMBHBs have similar variable mass transfer rates, then even those
systems that ought to be faint and persistent by virtue of their mean
$\dot M_T$ could be driven into transient behavior by fluctuations in
$\dot M_T$. Note also that the existence of faint persistent LMBHBs is
unlikely if the transition radii are substantially smaller than the
observational upper limits inferred in quiescent BH SXTs, or if
$\alpha_{\rm ADAF}$ is smaller than $0.3$ (e.g. $0.1$).

Currently, there is a discrepancy between theoretical estimates of the
rate of formation of black hole binaries and observational estimates of
the number of LMBHBs in the Galaxy; the former falls short by a
significant factor (e.g. Portegies Zwart, Verbunt \& Ergma 1997).  If
in addition to the known population of SXTs there were also an
undiscovered population of faint persistent LMBHBs, the discrepancy
would clearly increase.

We have not considered Low Mass Neutron Star Binaries for several
reasons.  First, the energy advected via the ADAF onto a NS would be
reradiated from the NS surface (Narayan, Garcia \& McClintock 1997);
the reemitted radiation would modify the structure of the ADAF
(Narayan \& Yi, 1995) and, to some extent, the stability properties of
the thin disk as well (van Paradijs 1996).  Second, given the
uncertainties in the structure and emission spectrum of the boundary
layer between the accretion flow and the NS surface, it is difficult
to make any quantitative predictions.  Finally, LMNSBs may exhibit a
strong `propeller' effect (e.g. Zhang, Yu \& Zhang 1998) which is
difficult to model.

On the other hand, work on faint persistent LMNSBs is well worth the
effort.  Because the advected energy is reradiated from the NS
surface, faint persistent LMNSBs are expected to be brighter than
faint persistent LMBHBs.  A population of faint persistent LMNSBs may
thus be easier to identify than their black hole counterparts.
Recurrence time scales in LMNSBs are also shorter than in LMBHBs.
Therefore, one could more quickly determine whether a candidate LMNSB
is persistent or transient.

\section{Conclusion}
In this paper, we have investigated the global stability of LMBHBs
assuming that the accretion flow around the black hole is composed of
two zones: an ADAF for radii less than a transition radius $R_{\rm
tr}$ and a thin accretion disk for radii beyond $R_{\rm tr}$.

The critical issue is the location of the transition radius.
Observations of $H_{\alpha}$ emission line widths constrain $R_{\rm
tr}$ in quiescent BH SXTs (\S 2.1.2).  The observational constraints are
consistent with an empirical prescription we have developed for
predicting $R_{\rm tr}$ on the basis of stream dynamics (\S 2.1.1) and
ADAF physics (\S 2.1.3).

The second issue is the mass transfer rates $\dot M_T$ in LMBHBs.  We
estimate $\dot M_T$ in the known SXTs from observations (\S 3.1.1 and 3.1.2)
and compare the estimates with the rates predicted by theoretical
models of binary evolution under the influence of gravitational
radiation, magnetic braking and secondary expansion (\S 4.1).  The
observational and theoretical estimates are in general agreement,
except in the case of magnetic braking where the predicted mass
transfer rate is much too large.  For this and other reasons
(explained in \S 4.1 and \S 4.2) we assume that the current prescriptions
for magnetic braking are invalid in BH SXTs and ignore magnetic
braking in our models.

By combining the theoretical and observational input on $R_{\rm tr}$
and $\dot M_T$, and comparing the derived values with the standard
model of the thermal-viscous instability in disks, we show that all
the known LMBHBs should be transients.  This is consistent with the
fact that they are all known to be SXTs, providing general
confirmation of our methods.

Our model also predicts that a number of LMBHBs, especially those
with somewhat longer orbital periods, might be non-transients.  These
steady systems would be faint and persistent, and extremely difficult
to discover.  The prediction of such a population of black hole
binaries is the main result of the paper.  We calculate model spectra
of these systems, and show that the spectra are very similar to those
of quiescent SXTs.  Unfortunately, we have been unable to come up with
an efficient search strategy to find these faint persistent LMBHBs.
We are also unable to estimate how large a population these systems
may constitute relative to the known population of BH SXTs.

\section*{Acknowledgments}

We are grateful to Vicky Kalogera, Uli Kolb and Jeff McClintock for
useful discussions and we thank Phil Charles, Jean-Marie Hameury and
Eliot Quataert for comments on the manuscript.  Insu Yi provided the
cooling function in Fig. 2a.  This work was supported in part by NASA
grant NAG 5-2837.  KM was supported by a SAO Predoctoral Fellowship
and a French Higher Education Ministry Grant.

\clearpage

\begin{table*}
\caption{ BH SXTs AND OUTBURST PARAMETERS}
\begin{center}
\begin{tabular}{lrcccccc} \hline \hline
\\
System & $P_{\rm orb}$ (hr)  & D (kpc) & i($^\circ$)& $m_1$ ($M_{\odot}$)& $m_2$
($M_{\odot}$) & $\Delta F$ (${\rm ergs \ 
cm^{-2}}$)& $\Delta t$ (yr)\\
(1)&(2)&(3)&(4)&(5)&(6)&(7)&(8)\\
\\
\hline
\\
GRO J0422+32 (XN Per. 92)&$5.1$ & $2.6^a$ & $30^g$ & $12^g$ & $ 0.4^e$& $1.2 \times
10^{-2}$& $> 30$\\
\\
A0620-00 (XN Mon. 75)&$7.8$ & $1^b$ & $55^b$ & $6.1^b$ & $0.55^e$& $3.3$& $58$\\
\\
GS2000+25 (XN Vul. 88)& $8.3$ & $2.7^b$ & $65^e$& $8.5^e$& $0.6^e$& $2.2$& $> 30$\\
\\
GS1124-683 (XN Mus. 91)&$10.4$ & $5^c$& $60^c$& $6^c$& $0.8^e$& $0.5$& $> 30$\\
\\
H1705-250 (XN Oph. 77)& $16.8$ & $8.6^b$& $70^e$& $4.9^e$& $0.7^e$& $0.14$& $> 30$\\
\\
GRO J1655-40 (XN Sco. 94)& $62.7$ & $3.2^d$ & $70^d$& $7^d$& $2.3^f$& $4 \times
10^{-2}$& $> 30$\\
\\
GS2023+338 (V404 Cyg) & $ 155.4$ & $3.5^b$ & $56^b$& $12^b$& $0.9^e$& $3.7$& $33$ \\
\\
\hline
\end{tabular}
\label{tab:systems}
\end{center}
NOTE. -- 
(1) All dynamically confirmed BH SXTs are listed here with the exception
of 4U 1543-47 which does not have adequate data.
(2) Orbital periods from \cite{van96}.
(3) Distances to the systems. 
(4) Inclinations.
(5) Black hole masses.
(6) Companion masses.
(7) Outburst fluences from \cite{cheshr97}, taking into account
the listed distances.
(8) Inter-outburst timescales, assuming that no outburst was missed 
(see \cite{cheshr97} for the reliability of the assumption that the coverage
of the X-ray sky was complete during the last 30 years).
(a) McClintock, private communication (1997).
(b) Narayan et al. (1997c).
(c) \cite{esietal97}.
(d) \cite{hamlas97}.
(e) \cite{cheshr97}.
(f) Orosz \& Bailyn (1997).
(g) Beekman et al. (1997).
\end{table*}
 
\clearpage 

\begin{table*}
\caption{ OBSERVATIONAL DATA USED FOR THE ADAF MODELS OF 
GRO J0422+32, GS2000+25 AND NOVA OPH. 77}
\begin{center}
\begin{tabular}{lccccc} \hline \hline
\\
System & $\lambda _{\rm opt} ($\AA$)$ & $A_{V}$ (mag) & \% from ``disk'' 
& optical $\nu F_{\nu}$ & X-ray $\nu F_{\nu}$\\
(1)&(2)&(3)&(4)&(5)&(6)\\
\\
\hline
\\
GRO J0422+32 (XN Per. 92)& $6300$ & $1.2$ & $30-60$ & $3.66^{+3.16}_{-2.2} \times
10^{-14}$ & $< 2.7 \times 10^{-14}$\\
\\
GS2000+25 (XN Vul. 88)& $6250$ & $5$ & $1-11$& $1.87^{+1.85}_{-1.59} \times
10^{-13}$ & $< 1.2 \times 10^{-13}$\\
\\
H1705-250 (XN Oph. 77)& $6250$ & $0.5$ & $63-72$ & $8.1^{+6.9}_{-3.41} \times
10^{-14}$ & $< 3 \times 10^{-13}$ \\
\\
\hline
 
\end{tabular}
\label{tab:3obs}
\end{center}
NOTE. -- 
(2) Optical wavelength of observation.
(3) Extinctions to the systems in the V band, from 
\cite{calgar96}, \cite{chaetal91}, \cite{grietal78}.
(4) Fraction of light at $\lambda_{\rm opt}$ that comes 
from the accretion flow, from \cite{filmat95}, \cite{haretal96},  
\cite{haretal97}.
(5) Net fluxes in ${\rm ergs \ s^{-1} \ cm^{-2}}$ at the wavelength 
$\lambda_{\rm opt}$.
The original fluxes are taken from \cite{filmat95}, \cite{haretal96}, 
\cite{filetal97}) and have been corrected for interstellar extinction and
contamination by the light from the secondary.  
(6) X-ray upper limits, in ${\rm ergs \ s^{-1} \ cm^{-2}}$, from 
\cite{garmac97}.\\
\end{table*}

\clearpage

\begin{table*}
\caption{ ACCRETION RATE ESTIMATES}
\begin{center}
\begin{tabular}{lccc} \hline \hline
\\
System & $\dot{M}_{\rm accum}$ (${\rm g \ s^{-1}}$) & 
$\dot{M}_{\rm ADAF}$ (${\rm g \ s^{-1}}$) & $\dot{M}_{\rm tot}$ (${\rm g \ s^{-1}}$) \\
(1)&(2)&(3)&(4)\\
\\
\hline
 
\\
GRO J0422+32 (XN Per. 92)& $< 8.3 \times 10^{14}$&
$1.5-2.2 \times 10^{15}$ & $1.5-4.5 \times 10^{15}$\\
\\
A0620-00 (XN Mon. 75)& $2.5 \times 10^{15}$&$2.1-3.3 \times 10^{15}$ & $3-8.7 \times
10^{15}$\\
\\
GS2000+25 (XN Vul. 88)& $< 1.2 \times 10^{16}$&$1.2-2.6 \times 10^{15}$ & $< 0.9-2.2
\times 10^{16}$\\
\\
GS1124-683 (XN Mus. 91)& $< 2.1 \times 10^{16}$&$1.9 \times 10^{15}$ & $< 1.5-3.6
\times 10^{16} $\\
\\
H1705-250 (XN Oph. 77)& $< 3.5 \times 10^{15}$&$2.7-6.8 \times 10^{15}$& $4.1-15.5
\times 10^{15}$\\
\\
GRO J1655-40 (XN Sco. 94)& $< 5.9 \times 10^{14}$&$1.3-2.9 \times 10^{16}$ & $0.9-4.4
\times 10^{16}$\\
\\
GS2023+338 (V404 Cyg) & $5.9 \times 10^{16}$&$2.2-3 \times 10^{16}$ & $5.4-13.3 \times
10^{16}$\\
\\
\hline
\end{tabular}
\label{tab:7mdot}
\end{center}
NOTE. -- 
(2) Assumes a radiative efficiency $\eta =0.1$ for accretion in outburst. 
(3) From the ADAF spectral models shown in Figs.~7 and 8.
(4) Sum of the two previous columns, with an additional uncertainty of a
factor of $1.5$ on the upper and lower limits.
\end{table*}

\clearpage

\clearpage

\clearpage

\clearpage

\clearpage

\clearpage

\clearpage

\clearpage

\clearpage

\clearpage

\clearpage

\end{document}